\documentclass[a4paper,11pt]{article}
\pdfoutput=1 

\usepackage{jheppub} 

\usepackage[T1]{fontenc} 
\usepackage{amsmath}
\usepackage{amssymb}
\usepackage{slashed}
\usepackage{color}

\usepackage{gensymb} 
\usepackage[parfill]{parskip}
\usepackage[utf8]{inputenc}
\usepackage{enumitem}
\usepackage{graphicx}
\usepackage[makeroom]{cancel}
\usepackage{dsfont}
\usepackage{mathtools}
\usepackage{commath}
\usepackage{comment}
\usepackage{multirow}
\usepackage{hhline}
\usepackage{makecell}
\usepackage{xcolor}
\usepackage[toc,page]{appendix}
\usepackage{placeins} 
\usepackage{flafter}  
\usepackage{cleveref}
\usepackage{soul}
\usepackage{dsfont}

\DeclareRobustCommand{\Sec}[1]{Sec.~\ref{#1}}

\DeclareRobustCommand{\Eq}[1]{Eq.~(\ref{#1})}
\DeclareRobustCommand{\Eqs}[2]{Eqs.~(\ref{#1}), (\ref{#2})}

\definecolor{red1}{cmyk}{0,1,1,0.3}
\definecolor{purp1}{cmyk}{0.3,1,0,0}
\definecolor{blue1}{rgb}{0.2,0.4,0.65}
\definecolor{green1}{rgb}{0.2,0.65,0.4}

\newcommand{\cO}{\mathcal{O}}
\newcommand{\cL}{\mathcal{L}}

\newcommand{\cM}{\mathcal{M}}

\newcommand{\cF}{\mathcal{F}}
\newcommand{\cC}{\mathcal{C}}
\newcommand{\cK}{\mathcal{K}}

\newcommand{\cPT}{\chi\mathrm{PT}}
\newcommand{\BW}{\mathrm{BW}}

\newcommand{\GeV}{\mathrm{GeV}}
\newcommand{\MeV}{\mathrm{MeV}}

\newcommand{\Tr}{\mathrm{Tr}}
\newcommand{\idet}{\mathds{1}}

\newcommand{\eg}{\textit{e.g.}}
\newcommand{\ie}{\textit{i.e.}}

\newcommand\eqc{\,,}
\newcommand\eqd{\,.}
\newcommand\diag{\mathrm{diag}}
\newcommand\phys{\mathrm{phys}}
\newcommand{\mang}{\beta_q}
\newcommand{\mangtil}{\widetilde{\beta}_q }
\newcommand\hhhline[1]{\hhline{#1}}

\usepackage{caption}
\usepackage{subcaption}
\newcommand{\dfigwidth}{0.35\linewidth}

\title{A covariant description of the interactions of axion-like particles and hadrons}

\author[a]{Reuven Balkin,}
\author[b]{Ta'el Coren,}
\author[b]{Yotam Soreq,}
\author[c]{Mike Williams}

\affiliation[a]{Department of Physics, University of California Santa Cruz and Santa Cruz Institute for Particle Physics, 1156 High St., Santa Cruz, CA 95064, USA}
\affiliation[b]{Physics Department, Technion -- Israel Institute of Technology, Haifa 3200003, Israel}
\affiliation[c]{Laboratory for Nuclear Science, Massachusetts Institute of Technology, Cambridge, MA, USA}

\emailAdd{reuven.b@campus.technion.ac.il}
\emailAdd{tael.coren@campus.technion.ac.il}
\emailAdd{soreqy@physics.technion.ac.il}
\emailAdd{mwill@mit.edu}

\abstract{
We present a covariant framework for analyzing the interactions and decay rates of axion-like particles~(ALPs) that couple to both gluons and quarks.
We identify combinations of couplings that are invariant under quark-field redefinitions, and use them to obtain physical expressions for the prominent decay rates of such ALPs, which are compared with previous calculations for scenarios where ALPs couple exclusively to quarks or to gluons.
Our framework can be used to obtain ALP decay rates for arbitrary ALP couplings to gluons and quarks across a broad range of ALP masses.
}

\begin{document} 

\maketitle
\flushbottom
 
\section{Introduction}
\label{sec:intro}
	
Axions and axion-like-particles~(ALPs) are hypothetical pseudoscalar particles that appear in many extensions of the standard model~(SM).
They can be associated with solutions to the strong CP problem~\cite{Peccei:1977hh,Peccei:1977ur,Weinberg:1977ma,Wilczek:1977pj}, the hierarchy problem~\cite{Graham:2015cka}, and may also serve as mediators to dark sectors~\cite{Nomura:2008ru,Freytsis:2010ne,Dolan:2014ska,Hochberg:2018rjs,Fitzpatrick:2023xks} or as viable dark matter candidates~\cite{Preskill:1982cy, Dine:1982ah, Abbott:1982af}.
In particular, ALPs with $\cO(\GeV)$ mass are predicted by the heavy QCD solution to the strong-CP problem~\cite{Fukuda:2015ana, Agrawal:2017eqm, Agrawal:2017ksf, Gaillard:2018xgk, Gherghetta:2020keg,Gupta:2020vxb, Gherghetta:2020ofz, Valenti:2022tsc}.
As pseudo-Nambu–Goldstone bosons~(pNGBs) associated with a spontaneously broken global symmetry, ALPs naturally acquire small masses and weak interactions suppressed by the symmetry-breaking scale.
For axion and ALP reviews, see~\cite{Marsh:2015xka,Graham:2015ouw,Hook:2018dlk,Irastorza:2018dyq,Agrawal:2021dbo}. 
 
Such GeV-scale ALPs have been the focus of extensive theoretical and experimental study~\cite{Dolan:2017osp,Alves:2017avw,Marciano:2016yhf,Jaeckel:2015jla,Dobrich:2015jyk,Izaguirre:2016dfi,Knapen:2016moh,Artamonov:2009sz,Bauer:2018uxu,Mariotti:2017vtv,CidVidal:2018blh,Aloni:2018vki,Aloni:2019ruo,Bauer:2020jbp,Bauer:2021wjo,Sakaki:2020mqb,Florez:2021zoo,Brdar:2020dpr,DallaValleGarcia:2023xhh,Kyselov:2025uez,Afik:2023mhj,Balkin:2021jdr,Blinov:2021say,Balkin:2023gya,Bai:2021gbm,Pybus:2023yex,Bai:2024lpq}, with their production and decay rates forming the basis of most phenomenological analyses. 
These rates are typically computed using different techniques depending on the ALP mass. 
At low masses ($m_a<1\,$GeV), chiral perturbation theory ($\chi$PT) provides reliable predictions for exclusive decay channels~\cite{Georgi:1986df}. 
At higher masses ($m_a \gtrsim 2-3\,$GeV), perturbative QCD~(pQCD) yields inclusive rate estimates.
However, calculations in the mass region between where these two techniques are valid have proven to be challenging. 

A first step towards calculating ALP rates in this intermediate mass region was taken in Ref.~\cite{Aloni:2018vki} using a data-driven approach in the scenario in which the ALP only couples to gluons. 
This method is based on $SU(3)_F$ flavor symmetry and utilizes $e^+e^-$ scattering data to account for the various unknown hadronic form factors.
A preliminary study of the ALP-quark couplings in this context was done in~\cite{Cheng:2021kjg}, focusing on a specific UV model in which the ALP-quark interaction is aligned with the quark couplings to the $Z$ boson.

The importance of basis invariance under quark-field redefinitions for physical observables such as ALP decay rates was emphasized in Ref.~\cite{Bauer:2021wjo}. 
That work showed how some earlier calculations of the $K^+ \to \pi^+ a$ decay omitted relevant contributions - an issue that becomes apparent when the computation is performed in a generic basis. 
The resulting dependence on arbitrary field-redefinition parameters highlights the need for basis-independent formulations, analogous to the role of gauge invariance in ensuring physical consistency. 
An initial step toward systematically implementing such invariance in ALP decay calculations was taken in Ref.~\cite{Ovchynnikov:2025gpx}, see also~\cite{Bai:2024lpq}.

In this work, we go beyond previous works such as~\cite{Aloni:2018vki} and develop a field-redefinition independent description of the interactions of ALPs with arbitrary quark and gluon couplings in the absence of the weak interactions.
We explicitly identify field-redefinition invariants, which control the physical processes. 
Our description allows us to estimate the ALP decay rates within $\chi$PT at low masses and by using the data-driven approach of Ref.~\cite{Aloni:2018vki} at higher masses.
Our resulting rates are explicitly independent of the basis and can be used to derive GeV-scale ALP phenomenology for any quark or gluon couplings. 

The rest of this paper is organized as follows.
In \cref{sec:model}, we introduce the ALP covariant framework and identify the chiral-rotation invariants. 
In \cref{sec:ALPmesonInteractions}, we derive the various low-mass ALP interactions within $\chi$PT, while in \cref{sec:chiPTExtened} we extend our results to the mass gap between the $\chi$PT and pQCD regions. 
The generic ALP decay rates are estimated in \cref{sec:ALPdecays}, and are shown explicitly for few benchmark models in \cref{sec:benchmarks}. 
We conclude in \cref{sec:summary}. 
Further details are given in \cref{A:constructing,A:Ta_kappa,A:Ta_limits,A:decay_VP,A:kappa_BW}.

\section{Chiral-Covariant ALP framework}
\label{sec:model}
	
This section presents a field-redefinition independent framework for studying the ALP and its interactions with SM particles. 
We start by studying ALP-parton interactions at the GeV scale.
Then, we embed the ALP into the chiral Lagrangian and match the two descriptions.  
 
\subsection{Partonic-level Lagrangian}
\label{sec:quark_lagr}
	
We start by writing the ALP effective Lagrangian at the $\cO(\GeV)$ scale, where the ALP is denoted as $a$ with mass $m_a$.
We consider interactions with the light quarks, $q=(u,d,s)$, and with the gluons.
The couplings to both light quarks and gluons at this scale can be generated at the quantum level due to interactions with heavier fields, and are therefore present in the IR even if they are absent at some high UV scale~\cite{Bauer:2020jbp}.

The effective Lagrangian is given by 
\begin{align}
	\label{eq:L_aq}
	\cL_{a,\mathrm{part}}
=&  -\bar{q} M q
	+\frac{\partial_\mu a}{f_a} \bar{q} \, c_q\, \gamma^\mu\gamma_5 q
	-c_g\frac{a}{f_a} \frac{\alpha_s}{8\pi} G^{\mu\nu}_c\widetilde{G}_{\mu\nu}^c 
	-c_\gamma\frac{a}{f_a} \frac{\alpha}{4\pi} F^{\mu\nu}\widetilde{F}_{\mu\nu}\eqc
\end{align}
with 
\begin{align}
    M\equiv& M_qe^{2i\gamma^5\mang \frac{a}{f_a}}\eqc 
\end{align}
where $M_q=\diag\left(m_u,m_d,m_s\right)$ is the SM quark mass matrix, and $\alpha\equiv e^2/4\pi$ and $\alpha_s\equiv g_s^2/4\pi$ are the EM and strong gauge coupling strengths, respectively. 
The EM and strong field-strength tensors are denoted by $F^{\mu\nu}$ and $G^{\mu\nu}_a$, with $\widetilde{F}_{\mu\nu}=\frac{1}{2} \epsilon_{\mu\nu\rho\sigma}F^{\rho\sigma}$ and $\widetilde{G}_{\mu\nu}^c=\frac{1}{2} \epsilon_{\mu\nu\rho\sigma}G^{c,\rho\sigma}$ being their duals.\footnote{We use the convention of $\epsilon^{0123}=+1$.}
The ALP decay constant is $f_a$, which has mass dimension one, and $c_g$, $c_\gamma$, $c_q$, and $\mang$ are all dimensionless coupling constants, with $c_q$ and $\mang$ being diagonal $3\times3$ matrices in quark-flavor space.
We do not consider flavor-violating couplings, as there are strong bounds on the decay rates in such scenarios, \eg~see~\cite{Bauer:2021mvw,MartinCamalich:2020dfe}.
Moreover, if the weak interactions are neglected, vector-like ALP-quark couplings of the form $({\partial_\mu a}/{f_a} )\bar{q} \, c^V_q\, \gamma^\mu q$ can be eliminated via a field redefinition which leaves the rest of the Lagrangian unchanged .
Therefore, in the absence of weak interactions, \Eq{eq:L_aq} is a generic starting point.
It is straightforward to add the ALP couplings to leptons, but for simplicity we omit these here.
	
It is important to note that $c_g$, $c_\gamma$, $c_q$, and $\mang$ are not physical parameters by themselves, as their values depend on the choice of basis.
We can perform a field redefinition in the form of an axial rotation, which shifts the values of the couplings while leaving any physical result invariant, see discussions in \eg~\cite{Bauer:2020jbp,Bauer:2021wjo}.
In particular, consider performing the field redefinition
\begin{align}
	\label{eq:trans_q}
	q'= e^{-i\kappa_q \frac{a}{f_a}\gamma_5}q\eqc
\end{align}
with $\kappa=\diag(\kappa_u,\kappa_d,\kappa_s)$ being the dimensionless quark rotation parameters.
As a result of this redefinition, the parameters of the Lagrangian transform as
\begin{align}
	\label{eq:trans_couplings}
	\mang\rightarrow \mang+\kappa_q\eqc \quad
	c_q\rightarrow c_q-\kappa_q\eqc\quad
	c_g\rightarrow c_g-\langle \kappa_q\rangle\eqc \quad
	c_\gamma\rightarrow c_\gamma-N_c \langle \kappa_q QQ\rangle\eqc
\end{align}
where $\langle \dots \rangle\equiv2\,\Tr(\dots)$ with the trace being performed in flavor space, the EM charge of the quarks is $Q\equiv \diag\left(2/3,-1/3,-1/3\right)$, and $N_c=3$ is the number of QCD colors. 
Therefore, it is clear that the parameters $c_g$, $c_\gamma$, $c_q$, and $\mang$ depend on the arbitrary quark-rotation parameters $\kappa$, and thus, cannot be physical. 

There are eight total parameters---$c_g$, $c_\gamma$, and three each in $c_q$ and in $\beta_q$---and three independent axial rotations.
Thus, there must be five independent invariants under the field redefinition of \cref{eq:trans_q}, which we identify as
\begin{align}
	\label{eq:cinvs}
	\mangtil\equiv\mang+c_q \eqc \qquad
	\widetilde{c}_g\equiv c_g-\langle c_q\rangle \eqc \qquad
	\widetilde{c}_\gamma\equiv c_\gamma-N_c\langle c_q QQ\rangle \eqd
\end{align}
Therefore, \textit{any physical observable must depend only on these invariants}. 
We further note that $\mang$ is ill-defined for $M_q=0$ and thus any dependence on $\mang+c_q$ must be proportional to $M_q$ such that it vanishes in the limit $M_q\to0$.

Any other invariant combination of couplings can be written as a linear combination of the terms in \Eq{eq:cinvs}.
In particular, for light ALPs with $m_a\ll m_u$ it is sometimes convenient to instead use the following linear combinations of these invariants,
\begin{align}
	\label{eq:betainvs}
	\bar{c}_q\equiv\mangtil\eqc\qquad 
    \bar{c}_g\equiv c_g+\langle \mang\rangle \eqc \qquad
	\bar{c}_\gamma\equiv c_\gamma+N_c\langle \mang QQ\rangle \eqd
\end{align}
%

\subsection{Embedding the ALP in the chiral Lagrangian}
\label{sec:embedALP}

Our next step is to embed the ALP into the chiral Lagrangian, where we follow~\cite{Gasser:1984gg,Bando:1984ej,Fujiwara:1984mp,Georgi:1986df} and~\cite{Callan:1969sn,Coleman:1969sm}.
The chiral Lagrangian, including the ALP, is given by
\begin{align}
	\cL_{\chi} 
=   \cL_{\rm kin} + \cL_{\rm mass} + \cL_V 
    + \cL_{\rm WZ} + \cL_{a\gamma\gamma} \eqc
\end{align}
where $\cL_{\rm kin}$, $\cL_{\rm mass}$, $\cL_V$, $\cL_{\rm WS}$, and $\cL_{a\gamma\gamma}$ are the kinetic, mass, vector-meson, Wess-Zumino~(WZ), and ALP-photon terms, respectively.
Below, we describe each of these components in detail and derive the ALP-meson interaction terms, summarized in \cref{tab:a_interactions}.
An in-depth construction of this Lagrangian is given in \cref{A:constructing}.

\begin{table}[t]
	\centering
	\scriptsize
	\begin{tabular}{|c|c|c|c|c|c|c|c|c|}
	\hline
	Term & kinetic mixing & mass mixing & $PPPP$ & $\gamma PP$ & $VPP$ & $PVV$ & $a\gamma\gamma$ & $\gamma PPP$ \\
	\hline
	$\cL_{\rm kin}$ & \checkmark &  & \checkmark & \checkmark &  &  &  &  \\
	\hline
	$\cL_{\rm mass}$ &  & \checkmark & \checkmark &  &  &  &  &  \\
	\hline
	$\cL_{V}$ &  &  & \checkmark & \checkmark & \checkmark &  &  &  \\
	\hline
	$\cL_{\rm WZ}$ &  &  &  &  &  & \checkmark & \checkmark & \checkmark \\
	\hline
	$\cL_{a\gamma\gamma}$ &  &  &  &  &  &  & \checkmark &  \\
	\hline
	\end{tabular}
	\caption{Interactions generated by various terms in $\cL_\chi$, where 
    $\gamma$ and $V$ represent interaction basis photons and vector mesons, respectively, 
    $P$ represents both the pseudoscalar mesons and the ALP (\eg{} $PPPP$ represents both a 4-meson vertex and an ALP--3-meson vertex), while $a$ represents the ALP specifically. 
    Both kinetic and mass mixings are also generated between the ALP and the SM pseudoscalar mesons.}
	\label{tab:a_interactions}
\end{table}	
	
The basic building block of the chiral Lagrangian is $\Sigma$, which is related to the pNGBs of the broken symmetry, the pseudoscalar mesons, via
\begin{align}
    \label{E:sig_def}
	\Sigma
    \equiv
    e^{\frac{2iP}{f_\pi}} \eqc
\end{align}
where $f_\pi\approx93\,\MeV$ is the pion decay constant, and 
\begin{align}
    \label{eq:P_def}
	P
=   \frac{1}{\sqrt{2}}
    \begin{pmatrix}
		\frac{\pi^0}{\sqrt{2}}+\frac{\eta}{\sqrt{3}}+\frac{\eta'}{\sqrt{6}}&\pi^+&K^+\\
		\pi^-&-\frac{\pi^0}{\sqrt{2}}+\frac{\eta}{\sqrt{3}}+\frac{\eta'}{\sqrt{6}}&K^0\\
		K^-&\bar{K}^0&-\frac{\eta}{\sqrt{3}}+\frac{2\eta'}{\sqrt{6}}
	\end{pmatrix}\eqd
\end{align}
The $\eta$ and the $\eta'$ are mass eigenstates, expressed as linear combinations of the mesons $\eta_8$ and $\eta_0$ corresponding to the $U(3)$ generators $T_8=\frac{1 }{\sqrt{12}}\diag(1,1,-2)$ and $\frac{1}{\sqrt{6}}\idet$, respectively. 
These mass eigenstates are given by
\begin{align}
	\label{E:eta_def}
    \eta
&=  \cos\theta_{\eta\eta'}\eta_8-\sin\theta_{\eta\eta'}\eta_0\eqc\\
    \label{E:etap_def}
	\eta'
&=  \sin\theta_{\eta\eta'}\eta_8+\cos\theta_{\eta\eta'}\eta_0\eqd
\end{align}
To simplify many resulting expressions, we take $\sin\theta_{\eta\eta'}\approx-1/3$.
More precise determinations of $\theta_{\eta\eta'}$ exist, but this approximation is sufficient for our purposes. 

The kinetic term is given by
\begin{align}
	\label{eq:lkin}
	&\cL_{\rm kin}
=   \frac{f_\pi^2}{8}
    \langle D_\mu\Sigma D^\mu\Sigma^\dagger\rangle\eqc
\end{align}
where the covariant derivative is 
\begin{align}
	\label{eq:DSigma}
	D^\mu\Sigma
    \equiv
    \partial^\mu\Sigma  +ie A^\mu[\Sigma,Q]
    +i\frac{\partial^\mu a}{f_a}\{\Sigma,c_q\} \, ,
\end{align}
with $[\cdot,\cdot]$ and $\{\cdot,\cdot\}$ denoting the commutator and anti-commutator, respectively. 
\Cref{eq:lkin} contains the meson kinetic terms and the leading-order~(LO) 4-meson derivative interactions.
It also induces ALP-meson kinetic mixing, as well as an ALP-3-meson derivative interaction. 

The mass term, $\cL_{\rm mass}$, is given by 	
\begin{align}
	\label{eq:lmass}
	\cL_{\rm mass}
=   \frac{f_\pi^2}{4}B_0
    \langle \Sigma M^\dagger+\Sigma^\dagger M\rangle
	-\frac{m_0^2}{2} \left\{\frac{f_\pi[\arg(\det\Sigma)+
    c_g a/f_a]}{\sqrt{6}}\right\}^2\eqd
\end{align}
The first term represents the contribution to the meson masses from the quark masses with $B_0=m_\pi^2/(m_u+m_d)$. 
The second term accounts for the contribution of the axial anomaly in QCD to $\eta_0$, the meson associated with the anomalous $U(1)_A$ symmetry, where
\begin{align}
    \frac{f_\pi \arg(\det\Sigma)}{\sqrt{6}} 
=   \frac{\langle P \rangle}{\sqrt{6}}
=   \eta_0
=   -\sin\theta_{\eta\eta'}\eta
    +\cos\theta_{\eta\eta'}\eta'\eqd
\end{align}
$\cL_{\rm mass}$ induces 4-meson interactions, as well as ALP-meson mass mixing and ALP-3-meson interactions.
The measured masses and widths of all mesons are taken from the Particle Data Group~(PDG)~\cite{ParticleDataGroup:2022pth}, but mass differences between the neutral and charged mesons are neglected (\eg{} we take $m_{\pi^\pm}=m_{\pi^0}$).
Setting
\begin{align}
	\frac{m_s}{m_u+m_d}=-\frac{1}{2}+\frac{m_\eta^2}{m_\pi^2}\eqc \qquad m_0^2=3m_\eta^2-3m_\pi^2\eqc
\end{align}
ensures that the $\eta$-$\eta'$ mixing angle corresponds to that given by \cref{E:eta_def,E:etap_def}, as well as the correct tree-level $\eta$ mass.

Within the LO chiral expansion, the $\eta'$ mass is predicted to be
$m_{\eta'}^2 = 4m_\eta^2-3m_{\pi}^2\approx (1071\,\MeV)^2$, which is larger by $\sim10\%$ compared to the observed $m_{\eta'} \approx 957\,$MeV.
This is due to corrections to the $\eta'$ mass coming from higher orders in the chiral expansion.
In our numerical calculations, we use the measured value of the $\eta'$ mass.
The fact that this value of $m_{\eta'}$ does not match the one predicted by the chiral Lagrangian introduces a small dependence on unphysical parameters; however, we verified that this dependence induces only $\cO(20\%)$ corrections for $\kappa\sim\cO(1)$ and $\mang=0$.

The interactions between the ALP and the vector fields, both mesons and photons, are contained within $\cL_V$, $\cL_{\rm WS}$, and $\cL_{a\gamma\gamma}$.
The vector Lagrangian is typically defined using the building blocks $\xi_L$ and $\xi_R$, which are set to $\xi^\dagger_L=\xi_R=e^{i\frac{P}{f_\pi}}$, and are combined into the $e$ symbol, defined as 
\begin{align}
	e_\mu 
	\equiv& 
	\frac{i}{2}(\xi_L D_\mu \xi_L^\dagger+\xi_R D_\mu \xi_R^\dagger)\eqc
\end{align}
with
\begin{align}
    \label{eq:covariant_xiLR}
	D_\mu\xi_{R/L} 
	\equiv   
	\partial_\mu\xi_{R/L}+ie A_\mu\xi_{R/L}Q
	\pm i\frac{\partial_\mu a}{f_a}\xi_{R/L}c_q\eqd 
\end{align}
Finally, the $e$ symbol is used to define the vector Lagrangian as
\begin{align}
	\label{eq:lV}
	&   \cL_{V} 
	=   
	f_\pi^2\left\langle{\left(g V_\mu-e_\mu\right)}^2\right\rangle\eqc 
\end{align}
with $g\approx\sqrt{12\pi}$~\cite{Fujiwara:1984mp} being the vector coupling constant, and $V^\mu$ the vector mesons,
\begin{align}
	\label{eq:V_def}
	V
	=   \frac{1}{\sqrt{2}}
	\begin{pmatrix}
		\frac{\rho^0}{\sqrt{2}}+\frac{\omega}{\sqrt{2}}&\rho^+&K^{*+}\\
		\rho^-&-\frac{\rho^0}{\sqrt{2}}+\frac{\omega}{\sqrt{2}}&K^{*0}\\
		K^{*-}&\bar{K}^{*0}&\phi
	\end{pmatrix}\eqd
\end{align}
Since $e_\mu = eQ A_\mu+...$, it is easy to see that $\cL_{V}$ leads to vector-meson photon mixing.
The physical vector meson states,
\begin{align}
	\label{eq:meson_photon_mix}
	V^\mu_\phys= V^\mu-\frac{e}{g}Q A^\mu+\mathcal{O}\left( \frac{e^2}{g^2}\right)\eqc
\end{align}
acquire a universal mass of $M_V^2=2f_\pi^2g^2$.
In our numerical calculations, we use the measured vector-meson masses instead.

The interactions of the ALP with two vectors are found in $\cL_{\rm WS}$ and $\cL_{a\gamma\gamma}$. 
The latter is simply the ALP-photon explicit interaction of the UV theory, 
\begin{align}
	\cL_{a\gamma\gamma} 
    = 
    -c_\gamma\frac{a}{f_a} \frac{\alpha}{4\pi} F^{\mu\nu}\widetilde{F}_{\mu\nu}\eqd
\end{align}
Meanwhile, $\cL_{\rm WS}$ contains the interactions coming from the Wess-Zumino terms~\cite{Witten:1983tw,Kaymakcalan:1983qq,Fujiwara:1984mp}, which are needed in order to match the chiral anomaly of the UV theory.
When combined with $\mathcal{L}_{a\gamma\gamma}$, we find
\begin{align}
	\label{eq:aVV_comp}
	\cL_{\rm WZ}+\cL_{a\gamma\gamma}
	=&  -\frac{\alpha }{4\pi f_a}
	\tilde{c}_\gamma aF^{\mu\nu}\widetilde{F}_{\mu\nu}
	-\frac{g^2 N_c}{16\pi^2 f_\pi}	
	\langle\widetilde{P}V^{\mu\nu}\widetilde{V}_{\mu\nu}\rangle\nonumber\\
	&   +i\frac{N_c}{12\pi^2}\frac{e}{f_\pi^3}\epsilon_{\mu\nu\rho\sigma}A^\mu 
	\langle Q\partial^\nu \widetilde{P}\partial^\rho \widetilde{P}\partial^\sigma \widetilde{P}\rangle\eqc
\end{align}
where $V^{\mu\nu}\equiv\partial^\mu V^\nu-\partial^\nu V^\mu$, $\widetilde{V}_{\mu\nu}\equiv\frac{1}{2} \epsilon_{\mu\nu\rho\sigma}V^{\rho\sigma}$ and $\widetilde{P} \equiv P+\frac{f_\pi}{f_a} c_q a$.
Of note is the last interaction term, which explicitly violates the vector meson dominance~(VMD) hypothesis~\cite{Sakurai:1960ju} by featuring an explicit interaction with photons; this deviation is introduced in order to match observations~\cite{Fujiwara:1984mp}.
The WZ interaction is notable for featuring an interaction of an odd number of pseudoscalars, accompanied by an epsilon tensor.
	
\subsection{ALP-meson mixing}
\label{sec:ALPmesonMixing}
	
The presence of the ALP in the kinetic and mass terms leads to mixing between the ALP and the flavor-neutral pseudoscalar mesons $\pi^0$, $\eta$, and $\eta'$ 
\begin{align}
    \cL
    \supset
    \frac{1}{2} K_{bc}\partial_\mu\phi_b\partial^\mu\phi_c
    -\frac{1}{2} m^2_{bc}\phi_b\phi_c\,,
\end{align}
with $b,c=a,\pi^0,\eta,\eta'$. 
The kinetic- and mass-mixing matrices are given by
\begin{align}
	\label{eq:mix_kai}
	K_{ai}
    =&  
    \langle T_i c_q\rangle\eqc \quad \\
	\label{eq:mix_mai}
	m_{ai}^2
    =&  
    -2B_0\langle T_i\mang   
    M_q\rangle+\frac{m_0^2}{6}c_g \langle T_i\rangle\eqc
\end{align}
respectively, where $i=\pi^0,\eta,\eta'$ and $T_i$ are the $U(3)$ matrices of the mesons. 
Isospin breaking leads to additional mass mixing between the neutral mesons,
\begin{align}
    m_{\pi\eta}^2
    =   
    -\sqrt{\frac{2}{3}}\delta_Im_\pi^2\eqc
    \qquad\qquad 
	m_{\pi\eta'}^2
    =   
    -\sqrt{\frac{1}{3}}\delta_Im_\pi^2\eqc
\end{align}
with $\delta_I=\frac{m_d-m_u}{m_u+m_d}\approx\frac{1}{3}$ being the isospin-breaking parameter.
The $\eta$ and $\eta'$ states do not mix by construction, see \cref{E:eta_def,E:etap_def} and subsequent discussion.

The interaction basis states defined in \cref{eq:P_def} can be written in terms of the physical mass eigenstates as
\begin{align}
	P_i
=   P_{\phys,i}+\frac{f_\pi}{f_a} h_{ai} a_\phys - \sum_{j\neq i}S_{ij}P_{\phys,j}\eqc
\end{align}
with
\begin{align}
	\label{eq:mix_hai}
	S_{ij}
=   \frac{m^2_{ij}}{m_i^2-m_j^2}\eqc \qquad
	h_{ai}
=   \frac{m^2_{ai}-m_a^2K_{ai}
    +\sum_{j\neq i}m^2_{ij}\frac{m^2_{aj}-m_a^2K_{aj}}{m_a^2-m_j^2}}{m_a^2-m_i^2} 
    \eqc
\end{align}
being the meson-meson mixing angle and the ALP-meson mixing angle~\cite{Aloni:2018vki}, respectively.
We can write this concisely as 
\begin{align}
	\label{eq:P_phys_mat}
	P
=   P_\phys+\frac{f_\pi}{f_a} T_a a_\phys+\Delta_{\pi\eta}\eqc
\end{align}
where $P_\phys$ is the matrix of the physical meson states, 
\begin{align}
    T_a
	\equiv 
    h_{a\pi}T_{\pi}+h_{a\eta}T_{\eta}+h_{a\eta'}T_{\eta'} 
\end{align}
is the effective $U(3)$ matrix of the ALP, and 
\begin{align}
	\Delta_{\pi\eta}
	\equiv
	-\sum_{i\neq j}T_i S_{ij}P_{\phys,j} 
\end{align}
accounts for SM mixing between the mesons, which is subleading and thus neglected in all relevant processes.

It is worth noting that $K_{ai}$ and $m_{ai}^2$, and thus $T_a$, are not invariant under the axial rotations.
Applying \cref{eq:trans_couplings}, we find  that  $T_a$ transforms as 
\begin{align}
	T_a\rightarrow T_a+\kappa_q\label{eq:trans_ta}\eqc
\end{align}
see \cref{A:Ta_kappa} for details. 
Since $T_a$ is not invariant, it is useful to define an invariant equivalent.
We define $\widetilde{T}_a\equiv T_a + c_q$ and use \cref{eq:mix_kai} and completeness relations to obtain 
\begin{align}
	\label{eq:ta_cq}
	\widetilde{T}_a
	\equiv T_a+c_q
=   \sum_i\frac{m^2_{ai}-m_i^2K_{ai}
    +\sum_{j\neq i}m^2_{ij}\frac{m^2_{aj}
    -m_a^2K_{aj}}{m_a^2-m_j^2}}
		{m_a^2-m_i^2}T_i\eqc
\end{align}
which is invariant under chiral transformations, see \cref{eq:trans_ta,eq:trans_couplings}, and thus measurable rates may depend on it. 
The quantity $\widetilde{T}_a$ is a function of the invariants defined in \cref{eq:cinvs}, namely $\widetilde{c_g}$ and $\mangtil$.
\Cref{eq:trans_ta,eq:ta_cq} are derived under the assumption of the LO chiral Lagrangian.
Therefore, using the measured physical meson masses  will introduce some basis dependency. 
Numerically, this is well below other theoretical uncertainties.

\begin{figure}[t]
  \centering
    \includegraphics[width=0.99\textwidth]
    {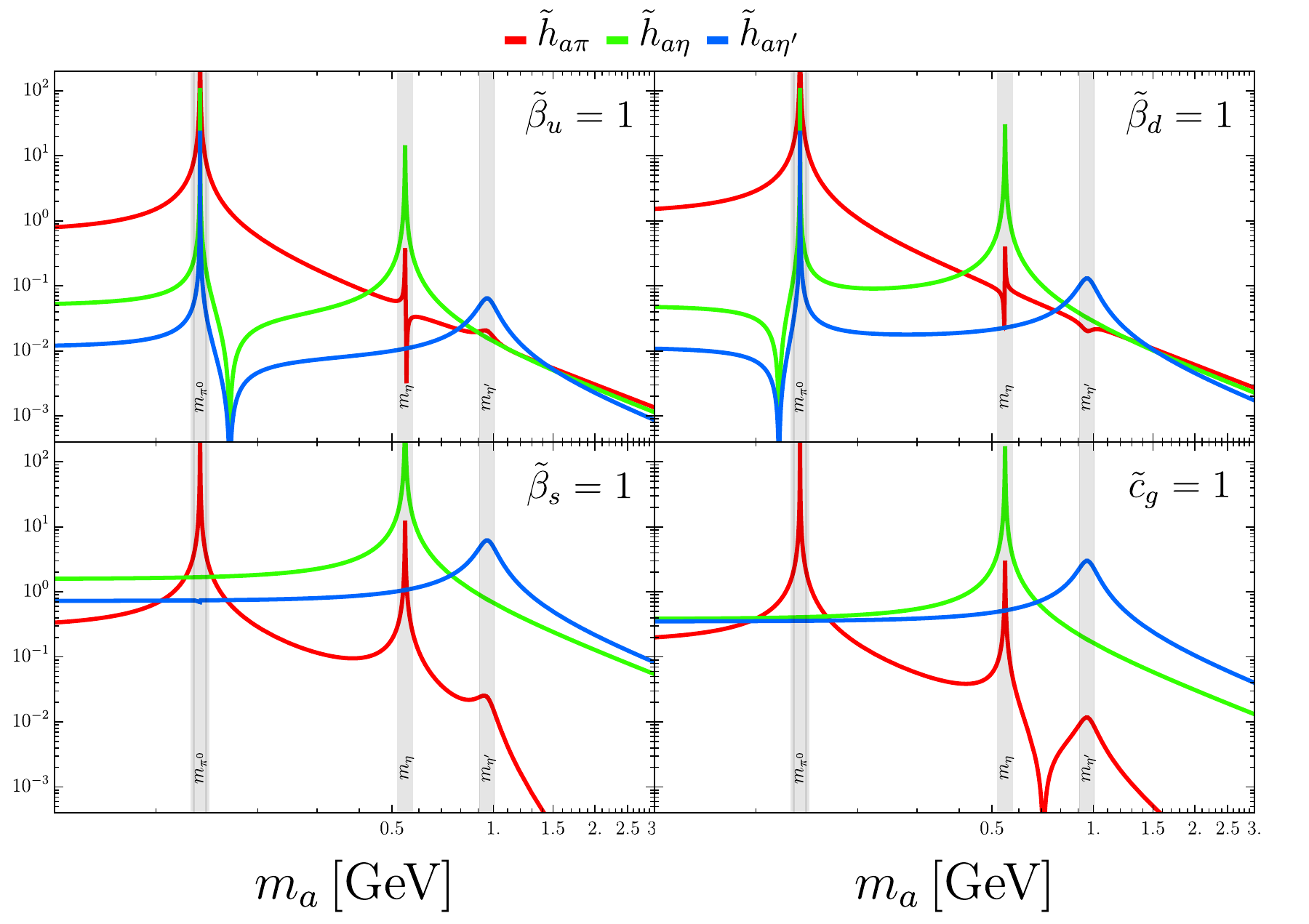}
      \caption{Dependence of $\tilde{h}_{a P}\equiv \langle \tilde{T}_a T_P \rangle$ on the four basis-independent coefficients $\tilde{\beta}_q = \text{Diag}[\tilde{\beta}_u,\tilde{\beta}_d,\tilde{\beta}_s]$ and $\tilde{c}_{g}$ as a function of ALP mass. In each panel, one of the four parameters is set to one, with the other 3 parameters set to zero.}
    \label{fig:Ta_coefs}
\end{figure}

In Fig.~\ref{fig:Ta_coefs} we show the dependence of $\tilde{h}_{a P}\equiv \langle \tilde{T}_a T_P \rangle$ on the four basis-independent coefficients $\tilde{\beta}_u,\tilde{\beta}_d,\tilde{\beta}_s$, and $\tilde{c}_{g}$ as a function of ALP mass. 
The effect of resonant mixing is clear as the ALP mass approaches one of the masses of the neutral mesons. 
Away from the resonant mixing regions, the couplings to $u$ and $d$ quarks typically make the ALP $\pi^0$-like, while the couplings to $s$ and gluons make the ALP $\eta$- or $\eta'$-like.

As stated above, at the light ALP limit, $m_a \ll m_\pi$, it is more convenient to use invariants involving $\mang$ instead of $c_q$, see \cref{eq:betainvs}. 
We find that
\begin{align}
    \bar{T}_a\equiv T_a-\mang
    &\xrightarrow{m_a \ll m_{\pi}}
    T_{a,0} \bar{c}_g \eqc
\end{align}
with
\begin{align}
    \label{eq:ta_lowmass}
    T_{a,0}
=&   -\frac{m_0^2}{3\sqrt{6}m_\eta^2}
    \left(T_\eta+\sqrt{\frac{2}{3}}\delta_IT_\pi\right)
    -\frac{2m_0^2}{3\sqrt{3}m_{\eta'}^2}
     \left(T_{\eta'}+\sqrt{\frac{1}{3}}\delta_IT_\pi\right) \eqc
\end{align}
a proof of which can be found in \cref{A:Ta_limits}.
For completeness, we find that $\widetilde{T}_a$ in this limit is $\widetilde{T}_a \xrightarrow{m_a \ll m_{\pi}} T_{a,0}\left(\widetilde{c}_g+\langle\mangtil\rangle\right)+\mangtil$.
%

\section{ALP-meson interactions}
\label{sec:ALPmesonInteractions}
 
The interactions of ALPs and hadrons can be roughly split into \textit{direct} and \textit{mixing} contributions.
The direct contributions originate from explicit ALP-meson vertices, \ie{} from the presence of $a(x)$ in the mass term and in the covariant derivative.
The mixing contributions originate from meson self-interactions, where one of the mesons is replaced by an ALP following \cref{eq:P_phys_mat}, \ie{} $P\to (f_\pi/f_a)T_a \, a$. 
However, since $T_a$ is not invariant under the chiral rotation it must always be accompanied by a corresponding direct contribution.
The mixing and direct contributions are combined into basis independent amplitudes, which  are functions of the invariants, \ie{} $\widetilde{T}_a$, $\mangtil$ and $\tilde{c}_\gamma$, with the latter only relevant being for the $a\rightarrow\gamma\gamma$ rate.

Terms which depend directly on $\mangtil$ are proportional to $M_q$ and are therefore typically subleading.
Thus, we find that most processes are a function of $\widetilde{T}_a$ alone.
We define the basis-independent combination
\begin{align}
	\label{eq:P_tilde_invariant}
	\widetilde{P}
	\equiv 
	P+\frac{f_\pi}{f_a} c_q a
	=
	P_\phys+\frac{f_\pi}{f_a} \widetilde{T}_a a_\phys+\Delta_{\pi\eta}\eqc
\end{align}
which shows up naturally in most vertices.
This is a generalization of Eq.~(6) in Ref.~\cite{Ovchynnikov:2025gpx} to any chiral basis and model. 

We derive the ALP-hadron interactions to leading order in the chiral expansion and in $f_\pi/f_a$.
A summary of all interactions is found in \cref{tab:a_interactions}.
	
\subsection{VPP}

Interactions between two pseudoscalars and a vector arise from two of the Lagrangian terms as follows.
The term $\cL_{\rm kin}$ in \cref{eq:lkin} is responsible for the interactions of photons with two pseudoscalars, while $\cL_V$ in \cref{eq:lV} contributes to pseudoscalar interactions with both photons and vector mesons.
Following the VMD paradigm (see \eg~\cite{Bando:1984ej}), the photon terms cancel and we are left only with the vector meson interactions
\begin{align}
    \label{eq:VPP}
	\cL_{VPP}
    =
    ig\langle V_\mu[\partial^\mu \widetilde{P},\widetilde{P}]\rangle
    -ig\langle \partial^\mu V_\mu[c_q\frac{f_\pi}{f_a}a, \widetilde{P}] \rangle\eqd
\end{align}
The first term is manifestly invariant under the field redefinition.
The the second term is not manifestly invariant, but it vanishes for an on-shell vector, $\partial_\mu V^\mu=0$. 
For processes involving vector exchange it can either vanish or cancel out other non-invariant contributions, such that the resulting amplitude is invariant under field redefinition; for example see \Sec{subsubsec:PPPP}.
	
\subsection{PVV}
\label{subsec:PVV}

The interaction of pseudoscalars with a pair of vectors is found in \cref{eq:aVV_comp} and consists of a $PVV$ vertex and an $a\gamma\gamma$ vertex.
Due to photon--vector-meson mixing, the $PVV$ vertex also contributes to the 
$PV\gamma$ and $P\gamma\gamma$ vertices in the mass basis.
Likewise, the $a\gamma\gamma$ vertex contributes to the $aV\gamma$ and $aVV$ vertices in the mass basis; this contribution is suppressed by a factor of $e^2/g^2$ and $e^4/g^4$, respectively. 
Since generally $\widetilde{T}_a\sim \widetilde{c}_g,\widetilde{c}_q$, we expect the $a\gamma\gamma$  contribution to $aV\gamma$\,($aVV$) to only be relevant when $\widetilde{c}_\gamma$ is larger by a factor of $g^2/e^2\approx 400$\,($g^4/e^4\approx 150000$) than the other couplings. We are interested in models where the ALP predominantly couples to gluons and quarks, thus,   models with large photon coupling are beyond our scope and we neglect the $a\gamma\gamma$ contribution to $aV\gamma$ and $aVV$.

To summarize, the vertices are 
\begin{align}
	&\cL_{PV_\phys V_\phys}=-\frac{g^2}{f_\pi}\frac{N_c}{16\pi^2} \langle\widetilde{P}V_\phys^{\mu\nu}\widetilde{V}^\phys_{\mu\nu}\rangle\eqc\label{eq:PVV}\\
	&\cL_{PV_\phys\gamma_\phys}=-\frac{eg}{f_\pi}\frac{N_c}{16\pi^2}\widetilde{F}_\phys^{\mu\nu}\langle\widetilde{P}\left\{Q,V^\phys_{\mu\nu}\right\}\rangle\label{eq:aVgam_phys}\eqc\\
	&\cL_{P\gamma_\phys\gamma_\phys}=-\frac{e^2}{f_\pi}\frac{1}{16\pi^2}\left(N_c\langle \tilde{P} QQ\rangle+\frac{f_\pi}{f_a} \tilde{c}_\gamma a\right)F_\phys^{\mu\nu}\widetilde{F}^\phys_{\mu\nu}\eqd
\end{align}
The last vertex gives an ALP-photon-photon interaction of
\begin{align}
	\cL_{a\gamma_\phys\gamma_\phys}
    =&
	-\frac{e^2}{f_a}\frac{\widetilde{\cC}_\gamma}{16\pi^2}a_\phys F_\phys^{\mu\nu}\widetilde{F}^\phys_{\mu\nu}\label{eq:agamgam_phys}\eqc
\end{align}
where $\widetilde{\cC}_\gamma\equiv\tilde{c}_\gamma+N_c\langle \tilde{T}_aQQ\rangle$.

\subsection{VPPP}

The chiral Lagrangian contains a $\gamma PPP$ vertex, see \cref{eq:aVV_comp}, given by
\begin{align}
    \label{eq:gammaPPP}
	&\cL_{\gamma PPP}=i\frac{N_c}{12\pi^2}\frac{e}{f_\pi^3}\epsilon_{\mu\nu\rho\sigma}A^\mu \langle Q\partial^\nu \widetilde{P}\partial^\rho \widetilde{P}\partial^\sigma \widetilde{P}\rangle\eqd
\end{align}
Unlike the $PVV$ and $VPP$ vertices, this vertex includes the photon rather than a vector meson, \ie{} VMD is not realized.
This term has been included, instead of a more general $VPPP$ vertex, to obtain better agreement with $\omega\rightarrow 3\pi$ measurements~\cite{Fujiwara:1984mp}.
As in \Sec{subsec:PVV}, photon--vector-meson mixing can give a contribution to a $V_\phys PPP$ vertex; however, this is highly suppressed.

\subsection{PPPP}
\label{subsubsec:PPPP}
	
There are multiple contributions to  4-pseudoscalar processes due to  terms originating from $\cL_{kin},\,\cL_{mass}$, and $\cL_V$.
Interactions in $\cL_{kin}$ and $\cL_{mass}$ contribute only as contact terms.
The relevant terms in the Lagrangian are given respectively by
\begin{align}
	\cL_{{\rm kin},4P}=&\frac{1}{6f_\pi^2}\left(\langle \left[\partial_\mu \widetilde{P},\widetilde{P}\right]^2\rangle-2\frac{f_\pi}{f_a} a\langle \left[c_q,P\right]\left[\partial^\mu\partial_\mu P,P\right]\rangle\right)\eqc\\
	\cL_{{\rm mass},4P}&=\frac{B_0}{3f_\pi^2}\left(\langle P^4M_q\rangle-4\frac{f_\pi}{f_a} a\langle P^3 M_q\mang\rangle\right)\eqd
\end{align}
Summing both contributions, we can bring the Lagrangian to the following almost manifestly invariant form: 
\begin{align}
    \label{eq:4P_kinmass}
	&\cL_{{\rm kin+mass},4P}
    =
    \frac{1}{6f_\pi^2}\langle \left[\partial_\mu \widetilde{P},\widetilde{P}\right]^2\rangle
    +\frac{B_0}{3f_\pi^2}\left(\langle \widetilde{P}^4M_q\rangle-4\frac{f_\pi}{f_a} a\langle \widetilde{P}^3 M_q\mangtil\rangle-\frac{f_\pi}{f_a} a \,\cO_{4P}\right)\eqd
\end{align}
The terms appearing explicitly in \cref{eq:4P_kinmass} are manifestly invariant, while the last term,
\begin{align}
  \cO_{4P} 
  \equiv  
  \langle \left[c_q,P\right]\left[\partial^2 P+B_0 \left\{P,M_q\right\},P\right]\rangle\,,
\end{align}
vanishes for on-shell mesons satisfying the equation of motion 
\begin{align}
    \partial^2 P=-B_0 \left\{P,M_q\right\}-\frac{m_0^2}{6} \langle P \rangle\mathds{1}\,.
\end{align}
The contact-term contribution from $\cL_V$ is given by
\begin{align}
    \label{eq:4P_V_naive}
	&\cL_{V,4P}=-\frac{1}{4f_\pi^2}\left(\langle \left[\partial_\mu \widetilde{P},\widetilde{P}\right]^2\rangle-2\frac{f_\pi}{f_a} a\langle \left[c_q,P\right]\left[\partial^\mu\partial_\mu P,P\right]\rangle\right)
    \eqc
\end{align}
where we note that $\cL_{V,4P}=-\frac{3}{2}\cL_{{\rm kin},4P}$.
Clearly, the first term in \cref{eq:4P_V_naive} is invariant and the second is not.
In order to get a physical result, \ie{} a 4-pseudoscalar scattering amplitude which depends only on the basis-independent combinations of \cref{eq:cinvs}, we must sum the contact term contributions of \cref{eq:4P_V_naive} and the factorizable contributions from vector exchange diagrams originating from the interactions in $\cL_{VPP}$, see \cref{eq:VPP}.
We recall that $\cL_{VPP}$ also contained both a manifestly invariant vertex, as well as a basis-dependent vertex proportional to $c_q$.
If we set the vector masses to their universal Lagrangian value $M_V^2=2g^2f_\pi^2$, we find that the basis-dependent contributions exactly cancel out, leaving us with a basis-independent result for the scattering amplitude, as required.     
In practice, we use the measured vector masses and widths, but still take the two unphysical contributions to cancel out exactly. 
	
Other sources of 4-pseudoscalar interactions are diagrams mediated by scalar resonances, as well as by the $f_2$ tensor meson. 
These are discussed in more depth in \cref{AA:scalars} and \cref{AA:tensor}, respectively.
	
\section{\texorpdfstring{Extending the chiral result to $1\,\GeV < m_a < 3\,\GeV$}{Extending the chiral result to 1 GeV < ma < 3 GeV}}
\label{sec:chiPTExtened}

The results obtained from the chiral Lagrangian are expected to be valid for $m_a \lesssim\GeV$.
In contrast, perturbative QCD becomes valid for $m_a\gtrsim 2\,\GeV$.
For the intermediate mass range of $1\,\GeV$ to $3\,\GeV$, we follow the data-driven approach of Ref.~\cite{Aloni:2018vki}.
We decompose the exclusive amplitudes with $Y$ external states denoted as 
\begin{align}
	\cM_Y=\cM_{Y,\cPT}\cdot\cF_Y\eqc
\end{align}
where 
$\cM_{Y,\cPT}$ is the amplitude calculated in chiral perturbation theory, and $\cF_Y$ is a hadronic form factor for the vertex, introduced to provide the corrections needed at higher masses. 
We posit that $\cF_Y$ only depends on the Lorentz representations of $Y$ (pseudoscalars, vectors, and baryons) and $\mu$, the hardest energy scale in the vertex, without depending on the flavor content of the particles involved or the softer momenta.
For simplicity, we take each $\cF_Y$ to be a real and positive function.
In the case of an amplitude with multiple vertices, we multiply each vertex by the relevant $\cF_Y$, evaluated at the highest energy scale for that vertex. 

The form factor for two vectors and a pseudoscalar, $\cF_{PVV}$, can be measured from $e^+e^-\rightarrow V^*\rightarrow VP$ processes~\cite{Aloni:2018vki}.
Fitting the $e^+e^-\to\rho\pi, \omega\pi,K^*K,\phi\eta$ data leads to
\begin{align}
    \label{E:F_PVV}
	&\cF_{PVV}(\mu)
    =
	\begin{cases}
		1, & \mu<1.4\,\GeV \\
		\text{interpolation}, & 1.4\,\GeV<\mu<2\,\GeV \\
		\left(\frac{1.4\, \GeV}{\mu}\right)^{X_{PVV}}, & 2\,\mathrm{GeV}<\mu
	\end{cases}\eqc
\end{align}
with $X_{PVV}=4$. 
In the low energy region $\cF_{PVV}$ is equal to one since chiral perturbation theory is valid.

The high-energy behavior of \cref{E:F_PVV} is in agreement with the theoretical expectations from Ref.~\cite{Lepage:1980fj}: 
A high-energy exclusive QCD process amplitude is expected to scale as $\cM_Y \propto \mu ^{4-n}$, with $n$ being the number of participating partons (incoming and outgoing quarks).
Since $\cM_{PVV,\cPT}$ by itself scales with two powers of the momentum, see \cref{eq:aVV_comp}, $X_{VVP}=4$ is needed to get the expected $\mu^{-2}$ behavior of a 3-meson process in agreement with the data. 

Regarding $\cF_Y$ for $Y\ne PVV$, in principle, one can use other data sets than $e^+e^-\to VP$ to measure $\cF_Y$ for the appropriate process. 
However, we take a different approach.
By following Ref.~\cite{Lepage:1980fj} we can find the high-energy scaling of any process and use it to estimate the $\cF_Y$ form factor.
For intermediate energies, we scale $\cF_Y$ using the measured $\cF_{PVV}$,
\begin{align}
    \label{E:F_gen}
	\cF_Y(\mu)=
	\begin{cases}
	   1, & \mu<1.4\,\GeV \\
		\cF_{PVV}\left(\mu\right)\left(\frac{1.4\,\GeV}{\mu}\right)^{\Delta X_Y}, & 1.4\,\GeV<\mu 
		\end{cases}\eqc
\end{align}
with $\Delta X_Y\equiv X_Y-X_{PVV}=X_Y-4$.
In the case that $\cF_Y(\mu)$ exceeds one it is set to one instead.
This form is consistent with the one given for $\cF_{PVV}$, has the correct high- and low-energy behavior, and is continuous everywhere.
The high-energy scaling $X_Y$ of all relevant vertices is summarized in \cref{T:scaling_powers}. 
This procedure was demonstrated, including comparison to data, for $\cF_{VPP}$ in Ref.~\cite{Aloni:2018vki}. 

In addition to the form factors, the high energy behavior of $\tilde{T}_a$ needs to be addressed.
In the limit of $m_a \gg m_{\eta'}$ \cref{eq:ta_cq} can be simplified to
\begin{align}
	\label{eq:Tatilhighmass}
	\widetilde{T}_a 
	\xrightarrow{m_a \gg m_{\eta'}}
	\frac{m_0^2\widetilde{c}_g\idet 
		-12B_0M_q \mangtil }{6m_a^2}+\cO\left(\frac{m_{\eta'}^4}{m_a^4}\right)\eqc
\end{align}
where a full proof is given in \cref{A:Ta_limits}.
The scaling $\tilde{T}_a \propto m_a^{-2}$ suppresses the rates at higher energies and is an artifact of the calculation in $\chi$PT. 
We note that in the case where the ALP couples solely to gluons $(\mangtil=0)$, equating the exclusive $\chi$PT rate to the inclusive pQCD rate leads to the matching condition $\widetilde{T}_a=\frac{\alpha_s(m_a)}{\sqrt{6}}\widetilde{c}_g\idet$~\cite{Aloni:2018vki}.\footnote{The normalization of $\widetilde{T}_a$ can be absorbed into $\cF$; thus, we adopt the normalization of Ref.~\cite{Aloni:2018vki}}
This can be interpreted as contributions to \cref{eq:Tatilhighmass} from heavier resonances.
We generalize this matching condition to the case of non-vanishing $\mangtil$, as we expect the same contributions to also affect the $M_q \mangtil$ term.
We set the relative coefficient between the terms by comparing the leading-order partonic $a\rightarrow qq$ and $a\rightarrow gg$ rates (given in \cref{subsec:partonic_rate}) in the high-energy region and demanding them to be equal for models with equal $\widetilde{T}_a$.
Putting this all together, for ALP masses above $1.2\,\GeV$, 
we use the matching condition
\begin{align}
	\label{eq:Tatilhighmass_corrected}
	\widetilde{T}_a 
	= 
	\frac{\alpha_s(m_a)}{\sqrt{6}}\widetilde{c_g}\idet -\sqrt{24}\pi\frac{M_q \mangtil }{m_a}\eqd
\end{align}
We note that naturally, as we switch from the chiral $\widetilde{T}_a$ to the high energy form, the ALP matrix is discontinuous at that point.
The crossover energy of $1.2\,\GeV$ was chosen to minimize this discontinuity, but does not eliminate it fully.
If one desires to have a continuous $\widetilde{T}_a$ they may use a smooth interpolating function in the crossover region.
In \cref{eq:Tatilhighmass_corrected}, we take  $\alpha_s$ from RunDec~\cite{Chetyrkin:2000yt} for $m_a>1.5\,\GeV$, and for $m_a\in [1.0,1.5]\,\GeV$ we use $\alpha_s(\mu) = \mu^{-2.6}$ for a smooth interpolation. 

\begin{table}[t]
	\centering
	\scriptsize
	\begin{tabular}{|c|c|c|}
		\hline
		Vertex & $\cM_{Y,\cPT}$ scaling & $X_Y$ \\
		\hline
		$PVV$ & 2 & 4 \\
		\hline
		$PV\gamma$ & 2 & 3 \\
		\hline
		$VPPP$ & 3 & 7 \\
		\hline
		$\gamma PPP$ & 3 & 6 \\
		\hline
		$VPP$ & 1 & 3 \\
		\hline
		$PPPP$ & 2 & 6 \\
		\hline
		$SPP$ & 2 & 4 \\
		\hline
		$TPP$ & 2 & 4 \\
		\hline
	\end{tabular}
	\caption{
    The high-energy power counting of the vertices considered in this work.
    }
    \label{T:scaling_powers}
\end{table}

\section{ALP decay rates}
\label{sec:ALPdecays}

In this section, we calculate the various ALP decay rates based on the framework presented in the previous sections. 
As we are attempting to estimate hadronic decay rates, we expect to have typical uncertainties of $20-30\,\%$ which can be even as large as $\cO(1)$ in some cases. 
For $m_a\lesssim 1\,\GeV$, we have better theoretical control over the predictions due to the chiral expansion, while for $m_a\gtrsim1\,\GeV$ we expect larger uncertainties.  
A dominant source for these theoretical uncertainties is the unknown strong phases of the different amplitudes, which in many cases can not be predicted  
from first principles.  
Additionally, for a heavier ALP, its decays receive contributions from heavier QCD resonances, \eg~\cite{BaBar:2021fkz}, which are not accounted for in our analysis, but could in principle be systematically added. 

Only searches which are looking for displaced signals are sensitive to the total decay rate of the ALP.
Current experiments are only capable of probing displaced ALPs in the sub-GeV mass region, where the expected error in the total rate is small. 
In addition, we note that for heavier ALPs, even an $\cO(1)$ error on various branching fractions will not have a significant affect on the predicted phenomenology. 

\subsection{$a\rightarrow VV$}

The decay of the ALP into a pair of vector mesons comes from the vertex detailed in \cref{eq:PVV}, and is found to be
\begin{align}
		\Gamma_{a\rightarrow V_1V_2}
        =
        \frac{N_c^2\alpha_g^2 m_a^3}{32\pi^3f_a^2S}
        \cF_{PVV}^2\left(\frac{1}{2} \langle \widetilde{T}_a\left\{T_{V_1},T_{V_2}\right\}\rangle\right)^2
        \left(1-\frac{4m_V^2}{m_a^2}\right)^{3/2}\eqc
\end{align}
where $\alpha_g\equiv\frac{g^2}{4\pi}\approx 3$, $T_{V_1}$ and $T_{V_2}$ are the $U(3)$ matrices of the two vector mesons, and $S=2$ for $V_1=V_2$ and one otherwise.
For all considered processes, the two final state particles have identical masses.
	
We note that due to the large width of the $\rho$ meson the narrow-width approximation is not valid, and it should instead be treated as a 4-body decay, involving two $\rho$-mediated diagrams.
For these processes, we use the 4-body phase-space integral from Ref.~\cite{Aloni:2018vki} and rescale it by the appropriate model-dependent factor $\langle \widetilde{T}_a \left\{T_\rho,T_\rho\right\}\rangle^2$.
In addition, we treat $a\rightarrow \rho\omega$ as a 3-body $a\rightarrow \pi\pi\omega$ decay, see below.
	
The decay rate into a vector meson and a photon is given by
\begin{align}
	\Gamma_{a\rightarrow V\gamma}
    =
    \frac{N_c^2\alpha\alpha_g m_a^3}{32\pi^3f_a^2}
    \cF_{PV\gamma}^2\left(\langle 
    \tilde{T}_aT_VQ\rangle\right)^2
    \left(1-\frac{m_V^2}{m_a^2}\right)^{3}\eqd
\end{align}
Due to flavor conservation, the vector meson must be flavor neutral.
As before, the narrow-width-approximation is not valid for the $\rho$, see below. 

The 2-photon decay rate is given by
\begin{align}
	\Gamma_{a\rightarrow \gamma\gamma}
    =
    \frac{\alpha^2 m_a^3}{64\pi^3f_a^2}\widetilde{\cC}^2_\gamma \eqd
\end{align}
Using \cref{eq:ta_lowmass} we find that
\begin{align}
    \widetilde{\cC}_\gamma(m_a\rightarrow0)
    \approx
    \bar{c}_\gamma-1.07 \, \bar{c}_g \, .
\end{align}
This result is in good agreement with existing calculations~\cite{GrillidiCortona:2015jxo}.
	
\subsection{$a\rightarrow VPP$}
\label{subsubsec:a_vpp}

\begin{figure}[t]
	\centering
	\begin{subfigure}[t]{\dfigwidth}
	   \includegraphics[width=\linewidth]{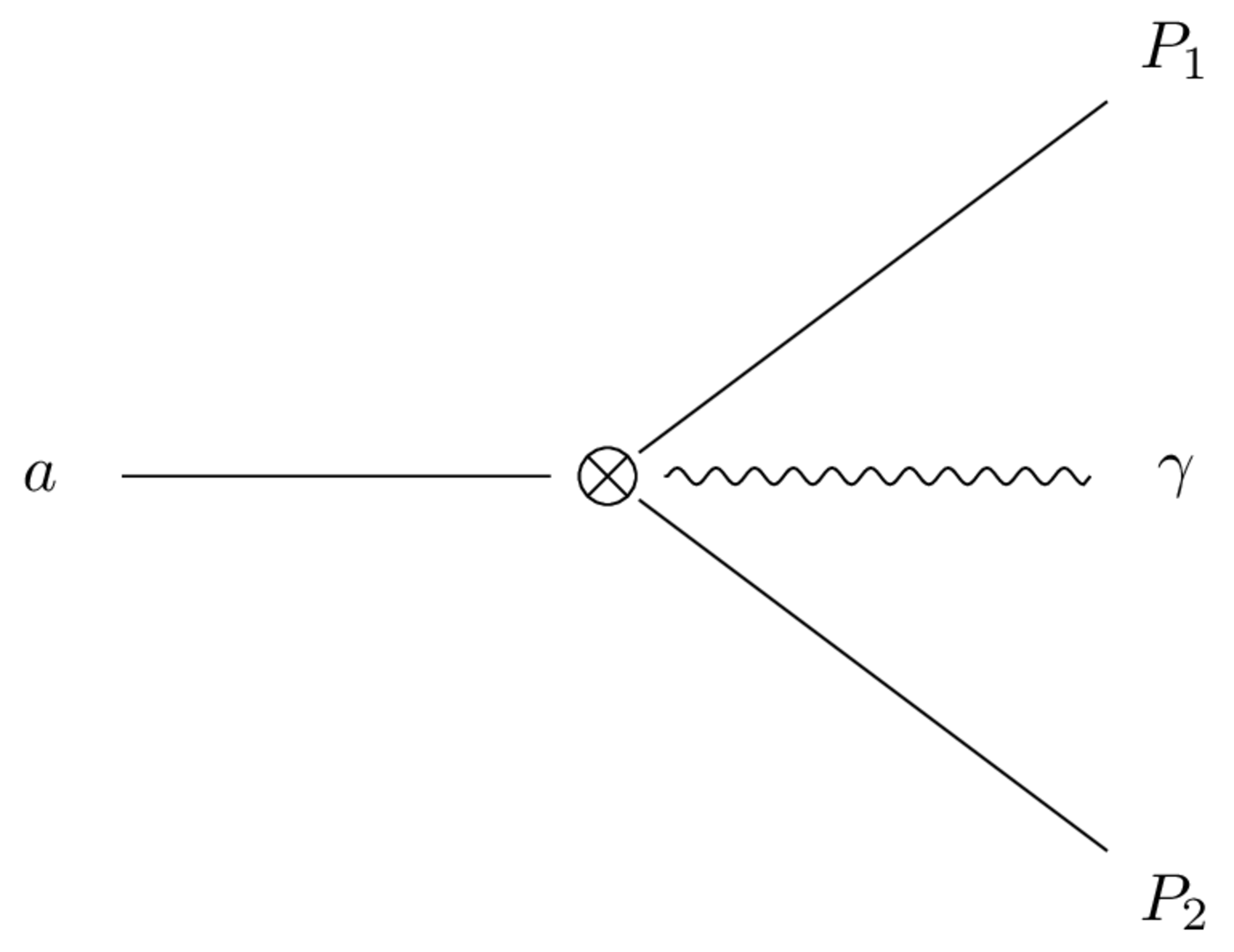}
	   \caption{Contact term.}
	\end{subfigure}
    \begin{subfigure}[t]{\dfigwidth}
		\includegraphics[width=\linewidth]{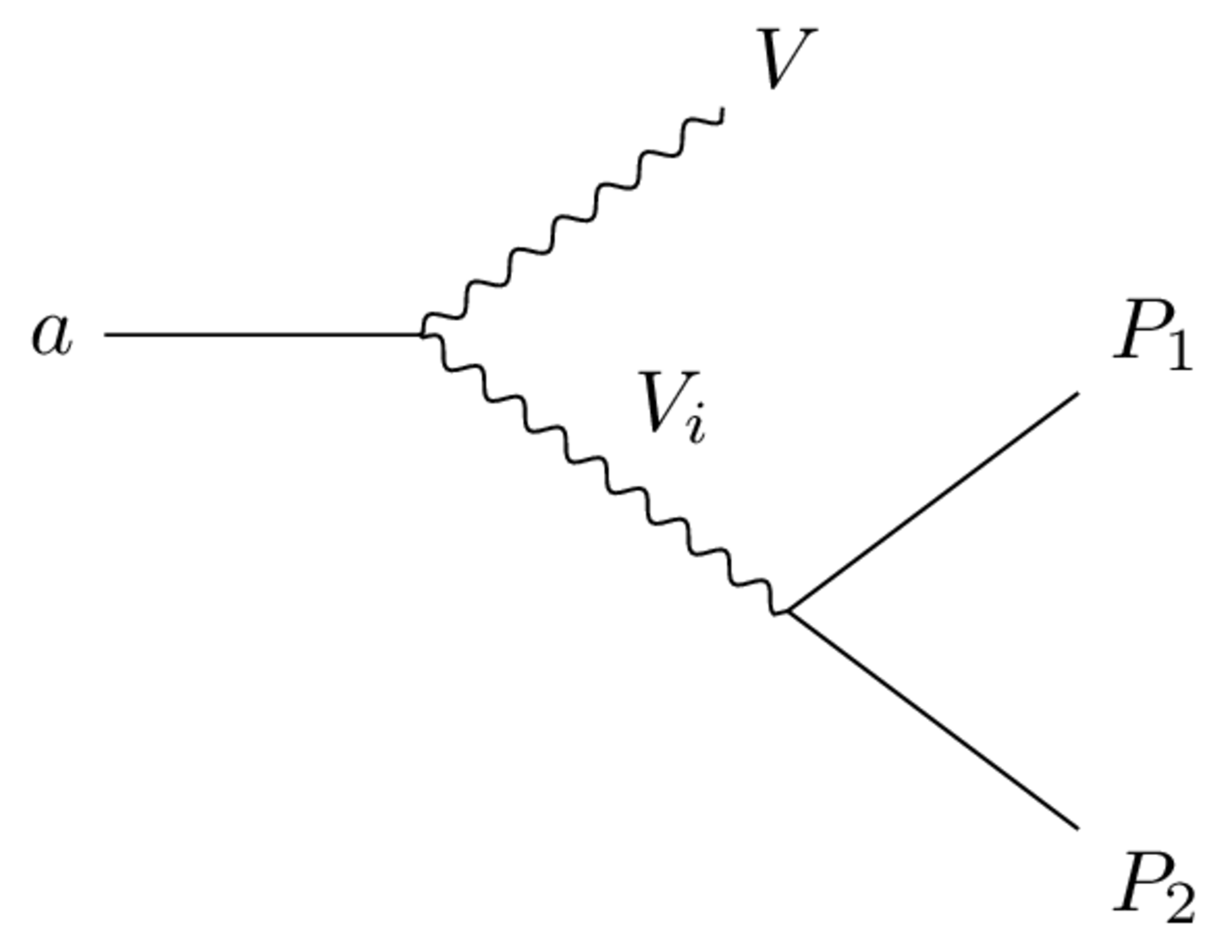}
		\caption{"$a$" type diagram.}
	\end{subfigure}	
    \begin{subfigure}[t]{\dfigwidth}
		\includegraphics[width=\linewidth]{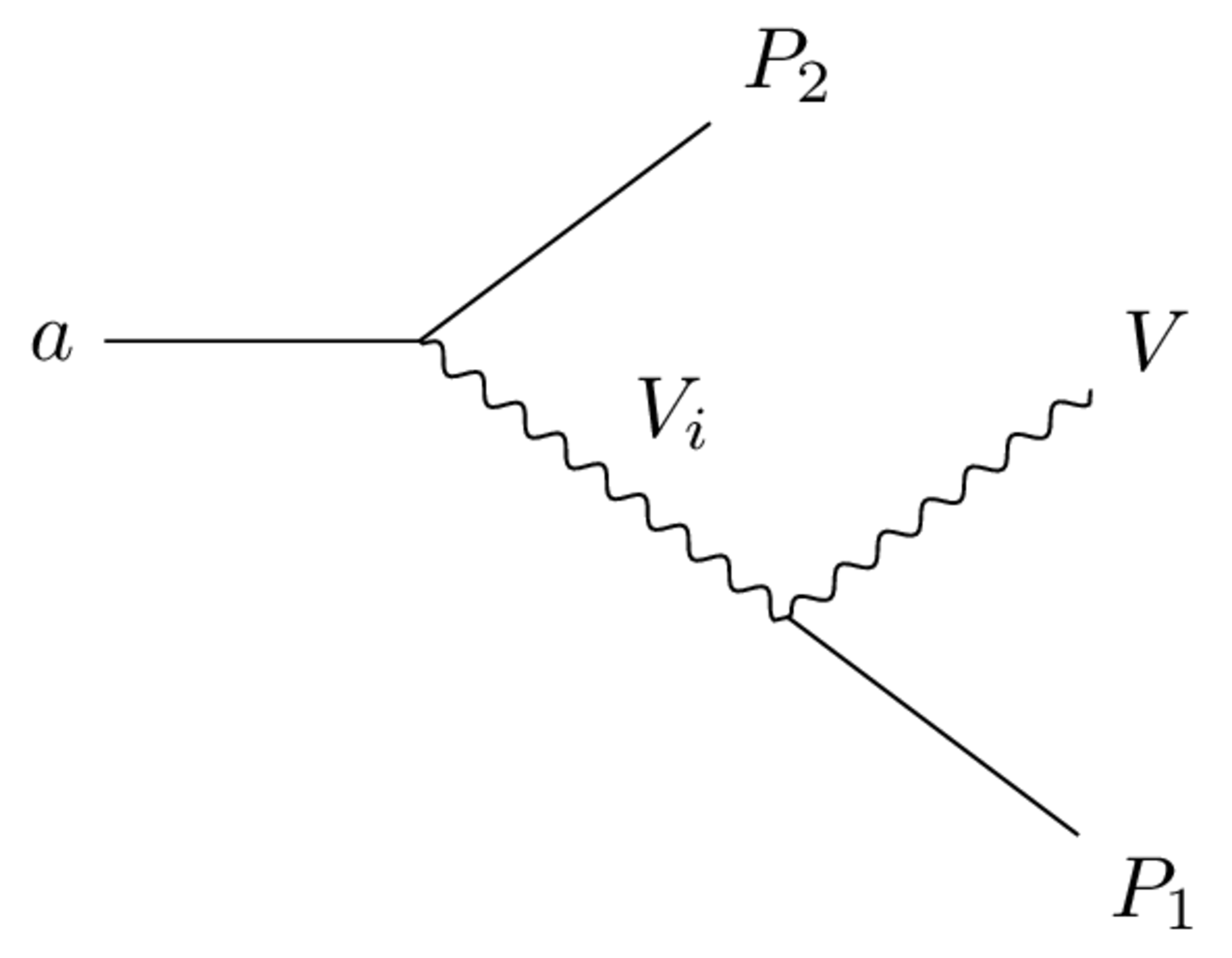}
		\caption{"1" type diagram.}
	\end{subfigure}
	\begin{subfigure}[t]{\dfigwidth}
		\includegraphics[width=\linewidth]{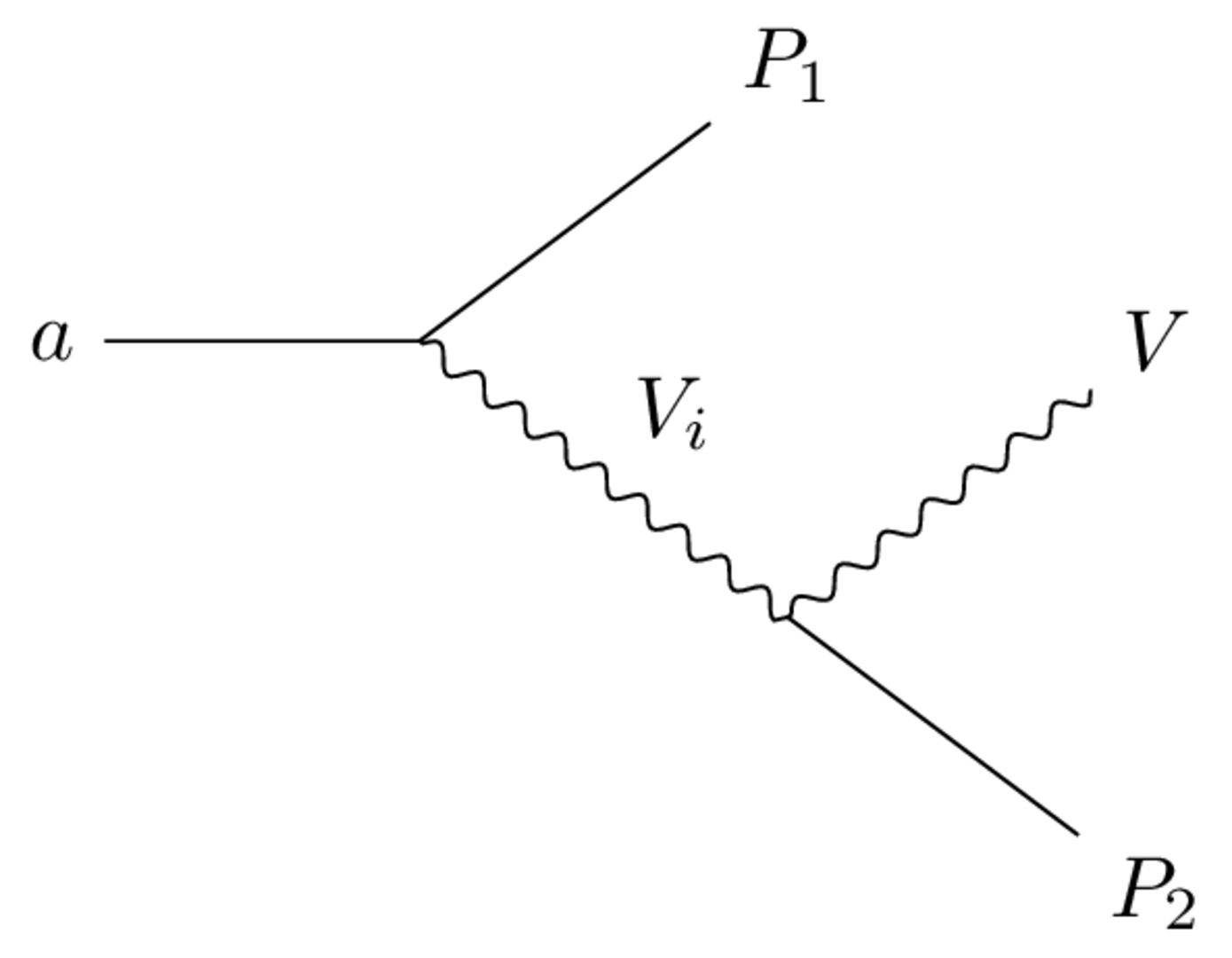}
		\caption{"$2$" type diagram.}
	\end{subfigure}
	\caption{Diagrams contributing to $a\rightarrow V P_1 P_2$. 
        The diagrams are labeled based on the pseudoscalar present in the $PVV$ vertex.}
	\label{fig:VPP_diags}		
\end{figure}

The decay of the ALP into a vector and two pseudoscalar mesons is facilitated by a combination of the $VPP$ and $PVV$ vertices in \cref{eq:PVV,eq:VPP}, as well as the direct $\gamma PPP$ contact term in \cref{eq:gammaPPP}.
The Feynman diagrams are shown in \cref{fig:VPP_diags}.
Summing over the four contributions leads to
\begin{align}
	&\cM_{a\rightarrow VP_1P_2} 
    =
    \frac{g^3N_c}{4\pi^2f_a}
    \left(A_{\mathrm{contact}}+A_a+A_1+A_2\right)
    \epsilon_{\mu\nu\rho\sigma}\epsilon^{*\mu}k_V^\nu k_1^\rho k_2^\sigma\eqc
\end{align}
where $\epsilon$ is the vector polarization, $k_V$,$k_1$,$k_2$ are the momenta of the final-state particles (1 and 2 here denote the two pseudoscalars). 
The $A$ terms, the labeling of which corresponds to the specific diagrams in \cref{fig:VPP_diags}, are
\begin{align}
    &A_a
    =
    +\cF_{PVV}\left(m_a\right)\cF_{VPP}\left(\sqrt{s_{12}}\right)
    \sum_{i}\langle T_i  \left\{T_V,\widetilde{T}_a\right\}
    \rangle \langle T_i^\dagger\left[T_1,T_2\right]\rangle 
    \BW_i\left(s_{12}\right)\eqc\\
	&A_1
    =
    -\cF_{PVV}\left(\sqrt{s_{1V}}\right)\cF_{VPP}\left(m_{a}\right)
    \sum_{i}\langle T_i^\dagger \left\{T_V,T_1\right\}\rangle 
    \langle T_i\left[\widetilde{T}_a,T_2\right]\rangle 
    \BW_i\left(s_{1V}\right)\eqc\\
    &A_2
    =
    +\cF_{PVV}\left(\sqrt{s_{2V}}\right)\cF_{VPP}\left(m_{a}\right)
    \sum_{i}\langle T_i^\dagger  \left\{T_V,T_2\right\}\rangle \langle T_i\left[\widetilde{T}_a,T_1\right]\rangle \BW_i\left(s_{2V}\right)\eqc\\
	&A_{\mathrm{contact}}
    =\frac{1}{3g^2f_\pi^2}\cF_{\gamma PPP}\left(m_{a}\right)
    \begin{cases}
		\frac{e}{g}\langle Q\left[\widetilde{T_a},T_1,T_2\right]\rangle, & 
        V=\gamma \\
		0, & V\neq\gamma
	\end{cases}\eqc\\
\end{align}
where $T_V$, $T_1$, $T_2$ are the $U(3)$ matrices (for the case $V=\gamma$, we take $T_\gamma=\frac{e}{g}Q$ as per the VMD paradigm, in addition to using $\cF_{PV\gamma}$ in place of $\cF_{PVV}$). 
The notation $s_{ij}$ denotes two particle invariant masses, \ie{} $s_{ij}\equiv \left(p_i+p_j\right)^2$, with $s_{12}+s_{1v}+s_{2v}=m_a^2+m_V^2+m_1^2+m_2^2$.
Additionally, we define $[T_A,T_B,T_C]\equiv T_AT_BT_C+T_BT_CT_A+T_CT_AT_B-T_AT_CT_B-T_CT_BT_A-T_BT_AT_C$. 
Finally, the sum runs over all mediating vectors, denoted by $i$, and the propagator is defined as
\begin{align}
    \label{eq:BW_def}
    \BW_j\left(p^2\right)
    \equiv 
    \frac{1}{p^2-m^2_j+im_j\Gamma_j}\eqd
\end{align}
For the $\rho$ mesons, we instead use a modified, mass-dependent Breit-Wigner function, following Ref.~\cite{BaBar:2012bdw}. 
We note that $A_\mathrm{contact}$ explicitly breaks VMD, see discussion below \cref{eq:gammaPPP}.
The total decay rate is then found in the standard way by integrating over the 3-body Lorentz-invariant phase space.

The $a\rightarrow VPP$ decays which are of the most interest to us are $a\rightarrow \gamma \pi^+\pi^-$ and $a\rightarrow \omega \pi^+\pi^-$, as they are found to be the most dominant. 
That being said, the outlined formulas may be used to find any $a\rightarrow VPP$ decay rate. 
However, care must be taken to avoid double counting these processes with processes such as $a\rightarrow K^*K^*$ followed by $K^*\to K\pi$ or $a\rightarrow V\phi$ followed by $\phi\to KK$.

\subsection{$a\rightarrow PPP$}

We can write the amplitude for a  $a\to PPP$ decay  as a sum of four contributions:
\begin{align}
   \mathcal{M}_{a\to PPP} = \frac{f_\pi}{f_a}\left[\mathcal{M}_{\text{mass}}+\mathcal{M}_{\text{kin}+V}+\mathcal{M}_{V}+\mathcal{M}_{s}\right]\,.
\end{align}
The first two contributions are contact terms arising from the manifestly basis-independent terms in \Eqs{eq:4P_kinmass}{eq:4P_V_naive}, which are given by
\begin{align}
	&\cM_{\rm mass}
    =   
    \frac{B_0}{3f^2_\pi }
    \left(\langle \widetilde{T_a}\left\{T_1,T_2,T_3,M_q\right\}\rangle
    -4\langle M_q\tilde{\beta}_q\left\{T_1,T_2,T_3\right\} \rangle\right)\cF_{PPPP}(m_a)\eqc\\
	&\cM_{{\rm kin}+V}
    =
    -\frac{1}{6f^2_\pi }\langle \widetilde{T_a}\left[T_1,\left[T_2,T_3\right]\right]\rangle
    \left(s_{12}-s_{13}\right)\cF_{PPPP}(m_a)+\text{permutations}\eqc
\end{align}
where $T_1$, $T_2$ and $T_3$ are the matrices of the 3 outgoing mesons.
The label ``+permutations" stands for summing over all 3 cyclic permutations of $(1,2,3)$, and we define $\left\{\cdot,\cdot,...,\cdot\right\}$ as the symmetric permutation of the matrices inside, \eg{} $\left\{A,B,C\right\}=ABC+BCA+CAB+ACB+BAC+CBA$.
	
The vector-induced amplitude $\cM_{V}$ combines contact terms from \Eq{eq:4P_V_naive} and the factorizable contributions using the vertices in \Eq{eq:VPP}, as discussed in \cref{subsubsec:PPPP}. 
The basis-dependent pieces cancel out, resulting in the following basis-independent expression
\begin{align}
	\cM_{V}
    =
    -g^2\sum_V 
    &\langle \widetilde{T}_a\left[T_1,T_V\right]\rangle\langle T_V^\dagger\left[T_2,T_3\right]\rangle
    \left(s_{12}-s_{13}-\frac{1}{m_V^2}(m_a^2-m_1^2)(m_2^2-m_3^2)\right)\nonumber\\
    &\times \BW_V(s_{23})\cF_{VPP}(m_a)\cF_{VPP}(\sqrt{s_{23}})
+\text{permutations}\eqc   
\end{align}
where we sum over all possible mediating vectors, and include all 3 cyclic permutations of $(1,2,3)$.
We note that the calculation of $a\to VP$ is effectively included in the above $a\to PPP$ derivation, see details in \cref{A:decay_VP}.
Since the vector mesons $V$ decay promptly, we consider only $a\to PPP$, which include a contributions from on-shell $V$ mesons.
	
Next, we have scalar-mediated diagrams contributing to the $a\rightarrow P_1P_2P_3$ rate via $a\rightarrow S(P_2P_3) P_1$. The associated amplitude is
\begin{align}
	\label{eq:4P_s_mediated}
    \cM_{s}
    =
    \sum_s&\gamma_{\bar{s},23}\gamma_{s,1a} (p_2\cdot p_3) (p_a\cdot p_1)\nonumber\\
    &\times \BW_s(s_{23})\cF_{SPP}(m_a)\cF_{SPP}(\sqrt{s_{23}})+\text{permutations}\eqd
\end{align}
The $\gamma_{s,ij}$ coefficients are the $SPP$ and $SPa$ coupling constants, covered in more depth in \cref{AA:scalars}, with $\bar{s}$ being the anti-particle of s.
Again, we sum over all mediating scalars $s$ and all 3 cyclic permutations of $(1,2,3)$.
We turn off the contribution from the $\sigma$ meson to this process above the $2m_K$ threshold, as using a simple BW distribution for those masses is known to violate unitarity. 
For the $\kappa$ meson, we replace the naive BW distribution with the empirical $K\pi$ S-wave amplitude, measured by the BaBar collaboration~\cite{BaBar:2015kii}.
This amplitude is known to have a large contribution from the $K_0^*(1430)$, and is not well described as a sum of BW terms.
The treatment of this BW is covered in more detail in \cref{A:kappa_BW}.

Lastly, we consider a contribution from the $f_2$ tensor to the $a\rightarrow \eta^{(\prime)} \pi \pi$ rates, covered in detail in \cref{AA:tensor}.
This contribution gives an amplitude of 
\begin{align}
	\cM_{f_2}&=-g_{f_2PP}^2\frac{f_\pi}{f_a}\langle \widetilde{T}_a T_{\eta^{(\prime)}}T_{f_2}\rangle \cF_{TPP}(m_a)\cF_{TPP}(\sqrt{s_{\pi\pi}}) \BW(s_{\pi\pi})A_{f_2}\eqc
\end{align}
where
\begin{align}
    A_{f_2}&\equiv \left( q\cdot p_{\eta^{(\prime)}} \right)^2-\frac{1}{3} q^2\left( m_{\eta^{(\prime)}}^2 -\frac{\left(k_{\pi\pi}\cdot p_{\eta^{(\prime)}}\right)^2}{s_{\pi\pi}} \right)\eqc
\end{align}
with $k_{\pi\pi}\equiv p_{\pi_1}+p_{\pi_2}$ and $q\equiv p_{\pi_1}-p_{\pi_2}$.

\subsection{Total decay rate}
\label{subsec:partonic_rate}

For low ALP masses we set the total width of the ALP to be the sum of all exclusive widths.
For high ALP masses, it is set to the inclusive rate from a partonic-level calculation.
The crossover energy at which we switch between the two is the mass at which the two rates are the closest.
If the two rates are equal at multiple masses, the higher mass is chosen.

The inclusive rate is given by the sum of the $a\rightarrow gg, \bar{q} q, \gamma \gamma$  rates, 
\begin{align}
    \Gamma_{a} 
    =
    \Gamma_{a\rightarrow \bar{q}q}^{\rm LO}\left(1+\Delta_{\bar{q}q}\right)
    +\Gamma_{a\rightarrow gg}^{\rm LO}\left(1+\Delta_{gg}\right)\, 
     +\Gamma_{a\rightarrow \gamma\gamma}^{\rm LO} \, .
\end{align}
The LO contributions are given by 
\begin{align}
	\Gamma_{a\rightarrow \bar{q}q}^{\rm LO}
	&=
	\frac{3m_a}{4\pi f_a^2}\langle\sqrt{1-{4M_q^2}/{m_a^2}}\,M_q^2\mangtil^2\rangle\eqc\\
	\Gamma_{a\rightarrow gg}^{\rm LO}
	&=
	\frac{\alpha_s^2m_a^3}{32\pi^3 f_a^2}\abs{\widetilde{\cC}_g}^2\eqc\\
	\Gamma_{a\rightarrow \gamma\gamma}^{\rm LO}
	&=
	\frac{\alpha_{EM}^2m_a^3}{64\pi^3 f_a^2}\abs{\widetilde{\cC}_\gamma^{\rm UV}}^2\eqc
\end{align}
where
\begin{align}
	\widetilde{\cC}_g
	\equiv
	\widetilde{c}_g+\langle\mangtil  \cK\left(\frac{4M_q^2}{m_a^2}\right)\rangle\eqc\qquad
    \widetilde{\cC}_\gamma^{\rm UV}
	\equiv
	\widetilde{c}_\gamma+N_c\langle\mangtil  \cK \left(\frac{4M_q^2}{m_a^2}\right)QQ\rangle\eqc
\end{align}
and the loop function is given by 
\begin{align}
    \cK(x)
	\equiv
	xf^2(x)\eqc \qquad
    f(x)
	\equiv
	\begin{cases}
		\arcsin{\frac{1}{\sqrt{x}}}, & 
        1\leq x \\
		\frac{\pi}{2}+\frac{i}{2}\log{\frac{1+\sqrt{1-x}}{1-\sqrt{1-x}}}, & x<1
	\end{cases}\eqd
\end{align}
The next-to-LO (NLO) corrections are taken from~\cite{Spira:1995rr,Djouadi:2005gj}: 
\begin{align}
	\Delta_{\bar{q}q}
	=
	5.67 \frac{\alpha_s}{\pi}
    \eqc
    \qquad\quad 
	\Delta_{gg}
	=
	\left(97-\frac{14}{3}N_F\right)\frac{\alpha_s}{4\pi}
    \, ,
\end{align}
where $N_F=3$ is the number of active flavors. 
These rates are manifestly invariant under field redefinition.
Note that $\cK$ is vanishing at the limit of $M_q\ll m_a$. 
Furthermore, we note that the partonic and chiral $a\rightarrow \gamma\gamma$ rates agree to order $\alpha_{\rm EM}^2$, with differences being suppressed by additional factors of $\alpha_s$, $\alpha_{\rm EM}$, or $M_q/m_a$.

\section{Example models}
\label{sec:benchmarks}

In this section, we consider three benchmark models and derive the dominant hadronic decay rates and branching fractions for each model.
A generalization of our results for any model is straightforward. 
We define the three benchmark models using the invariants of \Eq{eq:cinvs}. 
\begin{itemize}
	\item Model 1: \textit{gluon dominance}  
    \begin{align}
        \widetilde{c}_g=1\,, \qquad  \widetilde{c}_\gamma=\mangtil=0 \, , 
    \end{align}
    This model is identical to the one used \eg{}  in~\cite{Aloni:2018vki}.
    In terms of the coupling in \Eq{eq:L_aq}, this model is defined as $c_g=1$, $c_\gamma=c_q=\mang=0$.
    \item Model 2: \textit{dark pions}
    \begin{align}
        \widetilde{c}_g=1\,, \quad  
        \tilde{c}_\gamma=-\frac{2}{3} \,, \quad
        \mangtil=\diag\left(\frac{1}{2} , -\frac{1}{2} , -\frac{1}{2}\right) \,,
    \end{align}
    which is the dark pions models of Ref.~\cite{Cheng:2021kjg}.
    Equivalently, this model can be written as $c_g=\mang=0$, $c_\gamma=0$, 
    $c_q=\diag\left(\frac{1}{2} , -\frac{1}{2} , -\frac{1}{2}\right)$.
    \item Model 3: \textit{strange dominance} 
    \begin{align}
        \mangtil=\diag\left(0,0,1\right) \, , \qquad
        \widetilde{c}_g=\widetilde{c}_\gamma=0 \, .  
    \end{align}
    This can also be written as $c_g=2$, $c_\gamma=2/3$,  $c_q=\diag(0,0,1)$ and $\mang=0$. 
\end{itemize}

In models where $\widetilde{c}_g$ and all $\widetilde{\beta}_q$ are of the same order, the rates are controlled by $\widetilde{c}_g$ and $\widetilde{\beta}_s$. 
Therefore, we chose the above benchmarks to represent this generic case. 
In models where the dominant couplings are $\widetilde{\beta}_{u}$ and/or $\widetilde{\beta}_{d}$, the decay rates will be smaller, and there are larger errors (matching of the inclusive and exclusive rates) than in the above benchmarks. 

\begin{figure}[t!]
	\centering
		\includegraphics[width=0.95\textwidth]{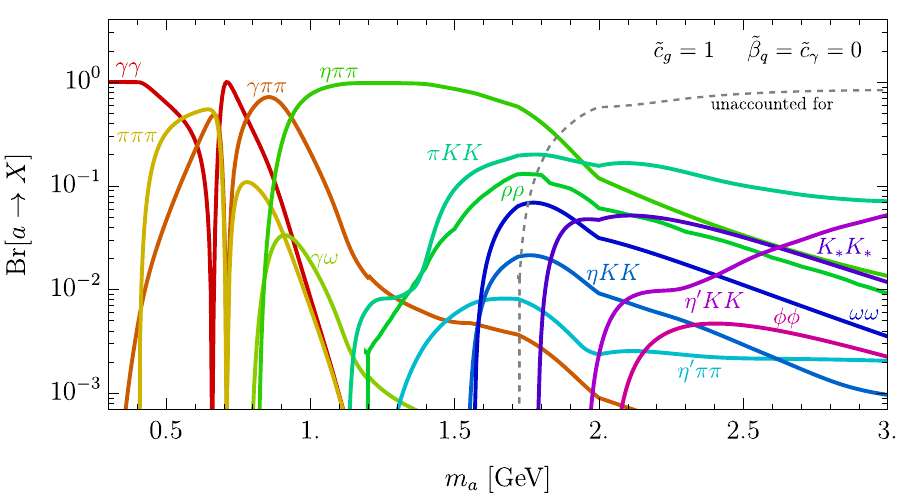}
        \\
\includegraphics[width=0.95\textwidth]{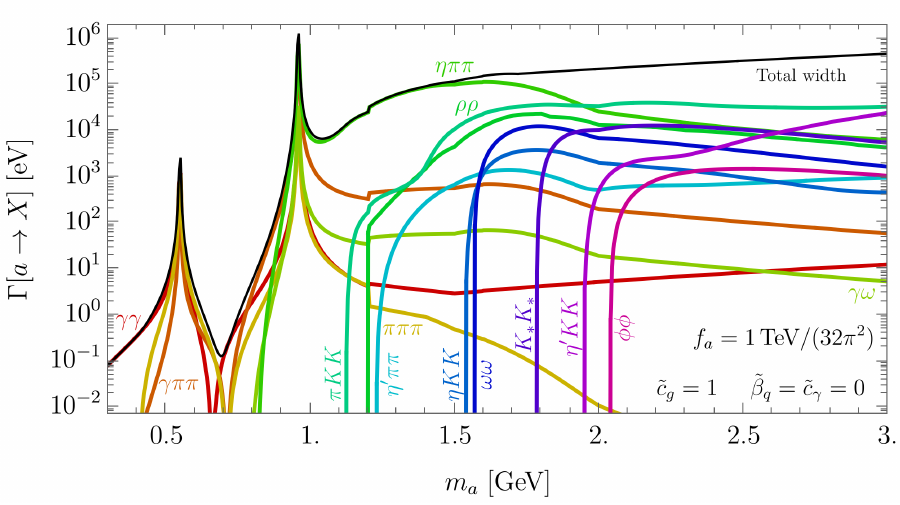}
	\caption{Model 1 ($\widetilde{c}_g=1$ and $\widetilde{c}_\gamma=\mangtil=0$): the hadronic branching fractions (top) and the partial decay widths (bottom). 
    The matching  between the inclusive and sum of exclusive modes is at $m_a=1.725\,\GeV$. The "unaccounted for" rate represents the difference between the sum of the exclusive rates and the inclusive rate.}
	\label{fig:model1}
\end{figure}

\begin{figure}[t]
	\centering		\includegraphics[width=0.95\textwidth]{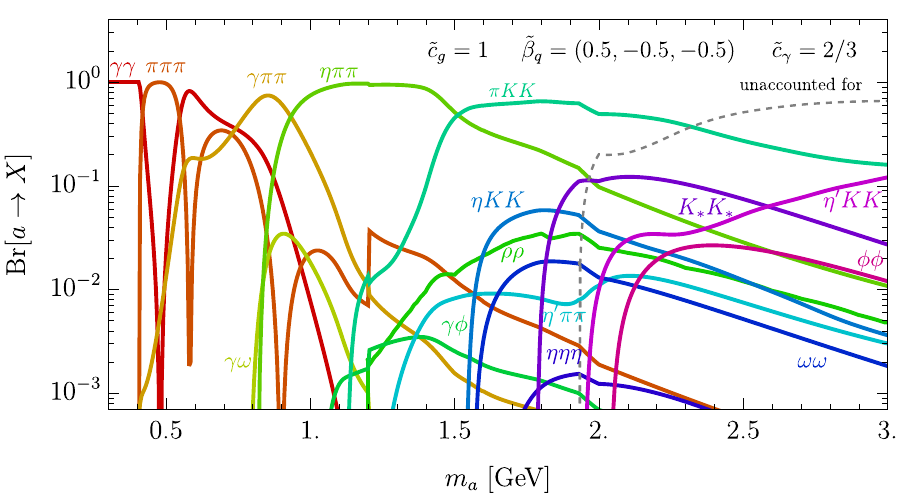}\\
\includegraphics[width=0.95\textwidth]{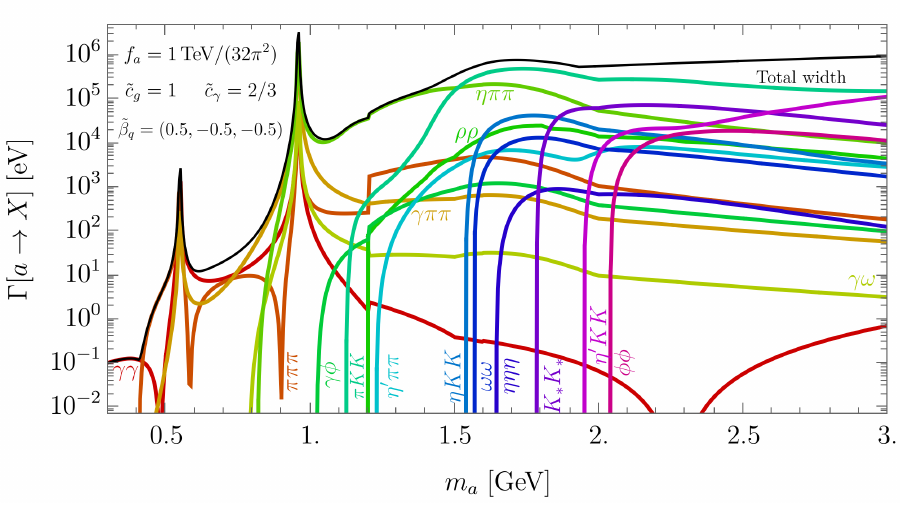}
	\caption{Model 2 ($\widetilde{c}_g=1$ and $\widetilde{c}_\gamma=-2/3$, $\mangtil=\diag(0.5,-0.5,-0.5)$): the hadronic branching fractions  (top) and the partial decay widths (bottom). 
    The matching  between the inclusive and sum of exclusive modes is at $m_a=1.935\,\GeV$. }
	\label{fig:model2}
\end{figure}

\begin{figure}[t]
	\centering	\includegraphics[width=0.95\textwidth]{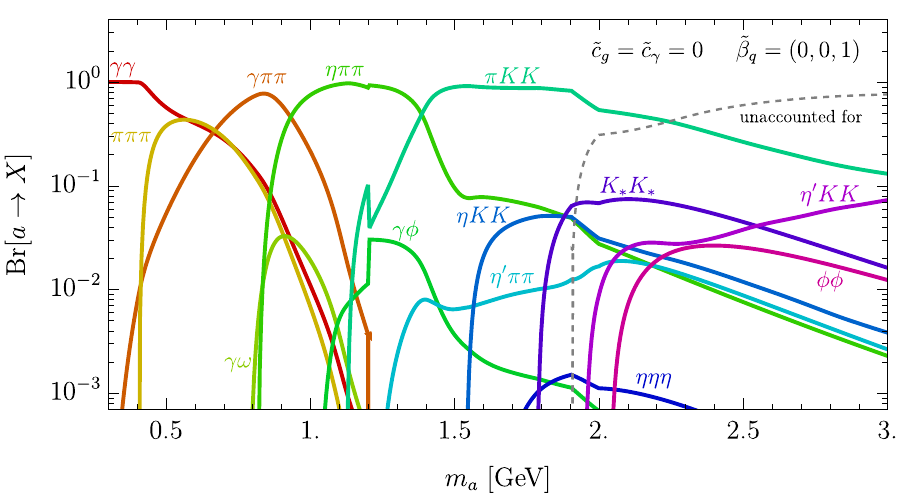}\\
\includegraphics[width=0.95\textwidth]{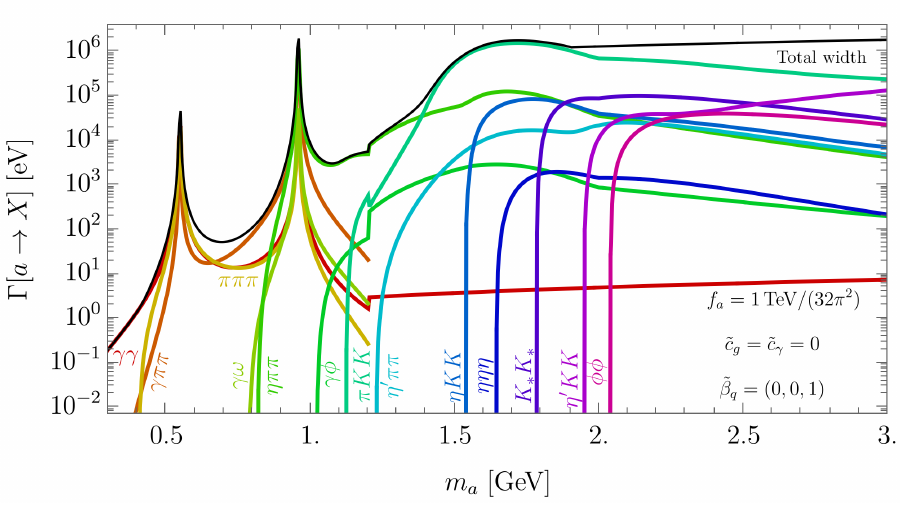}
	\caption{Model 3 ($\mangtil=\diag(0,0,1)$, $\widetilde{c}_g=\widetilde{c}_\gamma=0$): the hadronic branching fractions  (top) and the partial decay widths (bottom). 
    The matching  between the inclusive and sum of exclusive modes is at $m_a=1.91\,\GeV$. }
	\label{fig:model3}
\end{figure}

We show the hadronic branching fractions and the total hadronic decay width in \cref{fig:model1,fig:model2,fig:model3} for model 1,2 and 3, respectively. 
We find that the dominant decay mode for $m_a\lesssim 0.5\,\GeV$ is $\gamma\gamma$ as expected. For $0.5\,\GeV\lesssim  m_a \lesssim \GeV$, the $3\pi$ and $\pi\pi\gamma$ modes are dominant, while for $\GeV \lesssim  m_a < 1.5\,\GeV$ the dominant mode is $\eta\pi\pi$, and for models 2 and 3 it becomes $\pi KK$ for $m_a\gtrsim 1.5\,\GeV$.
We note that for model 1, the ALP's matrix becomes the identity at high energies, which forbids decays such as $a\rightarrow V(PP) P$. As such, observing such a decay can serve as proof for ALP-quark couplings, but may be experimentally challenging due to the large width the $\rho$ and the $K^*$ vector's proximity to the mass of the scalar $\kappa$ state.

%

\section{Summary}
\label{sec:summary}
 
In this work, we have presented a covariant framework for describing the interactions and decay rates of axion-like particles (ALPs) with arbitrary couplings to quarks and gluons in the absence of the weak interactions. 
A central feature of our approach is the explicit identification of five invariants under quark-field redefinition, which ensure that all physical observables---particularly ALP decay rates---are manifestly basis-independent. 
We construct the chiral Lagrangian including the interactions of pseudoscalars, vector mesons and scalars to leading order and derive expressions for decay amplitudes that consistently incorporate both direct and mixing contributions via the invariant matrix~$\widetilde{T}_a$.

Our framework reproduces known chiral perturbation theory results in the low-mass regime and extends them to higher masses using a generalized data-driven approach. 
We expand the results Ref.~\cite{Aloni:2018vki} to account for arbitrary quark and gluon couplings, update them by considering additional processes and diagrams, and overhaul the treatment of the $\kappa$ propagator.
By using the covariant approach, we ensured that all vertices are correct and physical.

We extended the framework of Ref.~\cite{Aloni:2018vki}, introducing appropriate hadronic form factors with scaling fixed by exclusive QCD power counting for all relevant vertices. 
To bridge the gap between the low- and high-mass regions, we construct a matching procedure for~$\widetilde{T}_a$, ensuring a smooth transition between hadronic and partonic descriptions.

We present numerical decay rate calculations for several benchmark models, highlighting how different couplings affect the dominant decay channels across the $0.1$--$3\,\mathrm{GeV}$ mass range. 
In particular, we find that gluon-dominated ALPs tend to decay via $\eta\pi\pi$ or $\gamma\pi\pi$ final states, while models with strange quark dominance or electroweak-aligned ALP couplings (\eg, the dark pion scenario of Ref.~\cite{Cheng:2021kjg}) can exhibit qualitatively distinct branching ratios.
Our results can be directly applied to compute ALP production and decay rates for arbitrary quark and gluon couplings.

\textit{Note added:} 
Ref.~\cite{Bai:2025fvl}, which has overlap with this manuscript, was posted when we where in final stages of this work.
	
\acknowledgments
We thank Daniel Aloni for collaboration during early stages of this work. 
TC, YS and MW are supported by the NSF-BSF (grant No. 2021800).
TC and YS are also supported by the ISF (grant No. 597/24).
MW is supported by NSF grant PHY-2209181.
RB is supported by the U.S. Department of Energy grant number DE-SC0010107. 
	
\appendix

\section{Constructing the ALP chiral Lagrangian}
\label{A:constructing}

In this appendix, we construct our low-energy theory by embedding the ALP into the SM chiral Lagrangian, term by term.

\subsection{External sources in ChiPT}
\label{AA:ALPspur}
Before embedding the ALP into the chiral Lagrangian in \cref{AA:embedALP} below, we recall that the chiral Lagrangian is based on the approximate $U(3)_L\times U(3)_R$ global symmetry of the SM. 
The most general starting point above the confinement scale is given by~\cite{Gasser:1984gg}
\begin{align}
	\label{eq:Lspur}
	&\cL=
	-\bar{q}M(x)q-\bar{q}\gamma_\mu(A_V^\mu(x)+A_A^\mu(x)\gamma^5)q-\theta(x)\frac{\alpha_s}{8\pi} G^{\mu\nu}_c\widetilde{G}_{\mu\nu}^c\eqc
\end{align} 
where $M(x),A_V^\mu(x),A_A^\mu(x)$ and $\theta(x)$ are external sources which could be in all generality  space-time dependent. 
We use the notation of the complex matrix $M$, \ie{} $M=Re(M)+iIm(M)\gamma_5$.
These sources may also explicitly break the chiral symmetry.
In order to properly integrate these potentially space-time dependent sources into the chiral Lagrangian below the confinment scale, we promote the global $U(3)_L\times U(3)_R$ symmetry to a local one~\cite{Gasser:1984gg}, under which the quarks transform as
\begin{align}
	\label{eq:trans_gen_q}
	&q\rightarrow \left(U_L(x)P_L+U_R(x)P_R\right)q\eqc
\end{align}
with $U_L$, $U_R$ being $3\times3$ matrices in flavor space and $P_{L}$ and $P_{R}$ being the left and right projection operators.
To ensure the invariance of \Eq{eq:Lspur} under this now local symmetry, we promote the external sources to spurions, transforming as
\begin{align}
	\label{eq:trans_gen}
	M\rightarrow U_L M U_R^\dagger\eqc\quad
	A_R^\mu \rightarrow U_R A_R^\mu U_R^\dagger +i U_R\partial^\mu U_R^\dagger\eqc\quad
	A_L^\mu \rightarrow U_L A_L^\mu U_L^\dagger +i U_L \partial^\mu U_L^\dagger\eqc
\end{align}
were we define $A_{R/L}^\mu\equiv A_V^\mu\pm A_A^\mu$. 
Since the transformation in \Eq{eq:trans_gen_q} is anomalous, $\theta(x)$ transforms by the trace of the infinitesimal axial transformation 
\begin{align}
	\label{eq:trans_theta}
	\theta \to \theta + \arg[\det (U_L^\dagger U_R)] \, \eqd
\end{align}
The embedding of any external source is then accomplished by requiring that the chiral Lagrangian be invariant under this now local symmetry.
Concretely, for the ALP model of \Eq{eq:L_aq}
\begin{align}
	\label{eq:gen_to_alp}
	A_A^\mu(x)=c_q\frac{\partial^\mu a}{f_a} \eqc\qquad
	M(x)=M_qe^{2i\gamma^5\mang \frac{a}{f_a}} \eqc\qquad		\theta(x)=c_g \frac{a}{f_a} \eqc
\end{align}
and $A_V^\mu$ is identified with the photon $A^\mu$, 
\begin{align}
	\label{eq:gen_to_alp_jv}
	&A_V^\mu(x)=eQA^\mu \eqd
\end{align}

\subsection{ALP chiral Lagrangian}
\label{AA:embedALP}

To construct the ALP chiral Lagrangian, we follow a simple procedure.  
We start with the SM theory, and demand that every term is invariant under the spuriunic $U(3)_L\times U(3)_R$ symmetry.
$\Sigma$, a fundamental building block of the chiral Lagrangian, transforms in the following way under $U(3)_L\times U(3)_R$,
\begin{align}
    \label{E:trans_sig}
	\Sigma\rightarrow U_L(x)\Sigma U_R^\dagger(x)\eqd
\end{align}

Starting with the kinetic term, we have
\begin{align}
	&\cL_{\rm kin}
	=   
    \frac{f_\pi^2}{8}
	\langle D_\mu\Sigma D^\mu\Sigma^\dagger\rangle\eqd
\end{align}
Demanding local $U(3)_L\times U(3)_R$ invariance, we find the covariant derivative to be
\begin{align}
	D^\mu\Sigma
	\equiv   
    \partial^\mu\Sigma
	+i\Sigma\left(A_V^\mu+A_A^\mu\right)-
	i\left(A_V^\mu-A_A^\mu\right)\Sigma\eqd
\end{align}
Since the covariant derivative contains $A_A^\mu$, and therefore the ALP, see \cref{eq:gen_to_alp}, the ALP is naturally embedded into the kinetic term, and \cref{eq:lkin,eq:DSigma} are derived.

For the mass term we have
\begin{align}
	\cL_{\rm mass}
	=   \frac{f_\pi^2}{4}B_0
	\langle \Sigma M^\dagger(x)+\Sigma^\dagger M(x)\rangle
	-\frac{m_0^2}{2} \left\{\frac{f_\pi[\arg(\det\Sigma)+\theta(x)]}{\sqrt{6}}\right\}^2\eqd
\end{align}
It is easy to see that the first term is invariant and reduces to the SM mass term in the absence of the ALP.
For the second term, we note that $\arg(\det\Sigma)$ and $\theta(x)$ are both invariant under $U(1)_V\times SU(3)_L \times SU(3)_R$, and shift with opposite signs under $U(1)_A$.
Making use of the presence of the ALP in $M(x)$ and $\theta(x)$, we derive \cref{eq:lmass}.

The vector-meson Lagrangian $\cL_V$ was derived using the Hidden Local Symmetry~(HLS) approach~\cite{Bando:1984ej}, which is based on the CCWZ construction~\cite{Coleman:1969sm}.
We introduce the fields $\xi_L$ and $\xi_R$, with $\Sigma = \xi_L^\dagger\xi_R$ and $\xi^\dagger_L=\xi_R=e^{i\frac{P}{f_\pi}}$.
The $\xi$ fields transform under the symmetry as
\begin{align}
	\label{E:xi_trans}
	\xi_L\rightarrow h\xi_L U_L^\dagger\eqc\qquad\qquad
	\xi_R\rightarrow h\xi_R U_R^\dagger\eqc
\end{align}
where $U_L,U_R$ are elements of the chiral group, while $h$ is a non-linear transformation under the $U(3)_V$ unbroken subgroup.
We note that $h$ is the unique $U(3)$ matrix such that the equality $\xi^\dagger_L=\xi_R$ is maintained under \cref{E:xi_trans}.
One then defines the $d$ and $e$ symbols used to construct the chiral Lagrangian as
\begin{align}
	\label{E:cov_xiR}
	e_\mu 
	\equiv& 
	\frac{i}{2}(\xi_L D_\mu \xi_L^\dagger+\xi_R D_\mu \xi_R^\dagger)  \;\;\;\; \text{and }\;\;\;\;
	d_\mu 
	\equiv 
	\frac{i}{2}(\xi_L D_\mu \xi_L^\dagger-\xi_R D_\mu \xi_R^\dagger) \eqc
\end{align}
where
\begin{align}
	D^\mu\xi_{L} 
	\equiv&   
	\partial^\mu\xi_{L}+i \xi_{L}A^\mu_V-i\xi_{L}A_A^\mu\eqc \\
	D^\mu\xi_{R} 
	\equiv&   
	\partial^\mu\xi_{R}+i \xi_{R}A^\mu_V+i\xi_{R}A_A^\mu\,.
\end{align}
The $d_\mu$ ($e_\mu$) symbol transforms (non-)homogeneously under $U(3)_V$, \ie{} as
\begin{align}
	d_\mu \to h  d_\mu h^\dagger\,, \;\;\;\;\; e_\mu \to h  (e_\mu+i\partial_\mu) h^\dagger\,.
\end{align}

The vector mesons are introduced as resonances transforming non-linearly as gauge bosons of $U(3)_V$, \ie{} with the same transformation properties as the $e$ symbol.
The unique invariant leading-order term is then given by~\cite{Bando:1984ej}
\begin{align}
	&   \cL_{V} 
	=   
	f_\pi^2\left\langle{\left(g V_\mu-e_\mu\right)}^2\right\rangle\eqd 
\end{align}

The construction of the Wess-Zumino term is more involved.
The term is split into the non-homogeneous terms, taken from Ref.~\cite{Kaymakcalan:1983qq}, as well as homogeneous terms, taken from Ref.~\cite{Fujiwara:1984mp}, which we shall tackle separately.

The Lagrangian given by Ref.~\cite{Kaymakcalan:1983qq} consists of multiple terms.
Each term is not individually invariant under the global symmetry, but the Lagrangian is constructed such that their sum total is invariant under non-anomalous transformations, thus fixing the relative coefficients of the different terms.  
They are given in \cref{tab:WZ_particular}, alongside their contributions to the to the $PVV$ and $VPPP$ vertices. 
We make use of the building block $\alpha\equiv \left(\partial_\mu \Sigma \right)\Sigma^\dagger dx^\mu$ therein, and the Lagrangian as a whole is proportional to the constant $C=\frac{-iN_c}{240\pi^2}$.
Of note is that we use Bardeen's form of the anomaly, which adds a counter term $-\Gamma_c$ to ensure the vector transformation (associated with the photon) remains anomaly free.
Summed together, they give
\begin{align}
	&\cL_{\rm WZ}^{ \text{\tiny non hom.}}=-\frac{N_ce^2}{16\pi^2f_\pi} F^{\mu\nu}\widetilde{F}^{\mu\nu} \langle PQ^2\rangle -\frac{iN_ce}{6\pi^2f_\pi^3}\epsilon_{\mu\nu\rho\sigma}A^\mu \langle Q\partial^\nu \widetilde{P}\partial^\rho \widetilde{P}\partial^\sigma \widetilde{P}\rangle\eqd\label{eq:WZ_particular}
\end{align}
We note that the unphysical (field-redefinition-dependent) terms in this expression cancel when it is combined with $\cL_{a\gamma\gamma}$.

\begin{table}[hbt]
	\centering
	\scriptsize
	\begin{tabular}{|c|c|c|}
		\hline
		Term & $PVV\cdot \frac{1}{5C}$ & $VPPP\cdot \frac{1}{40C}$ \\
		\hline
		$\frac{5}{2}iC\langle A_L\alpha^3\rangle-p.c.$ & $0$ & $\frac{e}{f_\pi^3}\epsilon_{\mu\nu\rho\sigma}A^\mu \langle Q\partial^\nu P\partial^\rho P\partial^\sigma P\rangle$ \\
		\hline
		$-\frac{5}{2}C\langle\left(dA_L A_L+A_LdA_L\right)\alpha\rangle-p.c.$ & $-2i\frac{e^2}{f_\pi}F^{\mu\nu}\widetilde{F}_{\mu\nu} \langle PQ^2 \rangle$ & $\frac{e}{f_\pi^3}\frac{f_\pi}{f_a}\epsilon_{\mu\nu\rho\sigma} A^\mu \langle Q c_q \partial^\nu a\partial^\rho P\partial^\sigma P \rangle$\\
		\hline
		$\frac{5}{2}C\langle d A_Ld\Sigma A_R\Sigma^\dagger\rangle-p.c.$& $-i\frac{e^2}{f_\pi}F^{\mu\nu}\widetilde{F}_{\mu\nu}\langle P Q^2\rangle$ & $\frac{1}{2}\frac{e}{f_\pi^3}\frac{f_\pi}{f_a}\epsilon_{\mu\nu\rho\sigma} A^\mu \langle Q\partial^\nu  P c_q \partial^\rho a \partial^\sigma P\rangle$\\
		\hline
		$-\frac{5}{2}C\langle A_L\Sigma A_R\Sigma^\dagger\alpha^2\rangle-p.c.$ & $0$ & $\frac{e}{f_\pi^3}\frac{f_\pi}{f_a}\epsilon_{\mu\nu\rho\sigma}A^\mu \langle Q  \partial^\nu P \partial^\rho P c_q \partial^\sigma a\rangle$\\
		\hline
		$\frac{5}{4}C\langle A_L\alpha A_L\alpha\rangle-p.c.$ & $0$ & $\frac{1}{2}\frac{e}{f_\pi^3}\frac{f_\pi}{f_a}\epsilon_{\mu\nu\rho\sigma}A^\mu \langle Q\partial^\nu P c_q \partial^\rho a\partial^\sigma P\rangle$\\
		\hline
		$\frac{5}{2}iC\langle\left(dA_R A_R+A_RdA_R\right)\Sigma^\dagger A_L\Sigma\rangle-p.c.$ & $-2ie^2\frac{a}{f_a}F^{\mu\nu}\widetilde{F}_{\mu\nu}  \langle c_q   Q^2\rangle$ & $0$\\
		\hline
		$-\Gamma_c$& $+2ie^2\frac{a}{f_a}F^{\mu\nu}\widetilde{F}_{\mu\nu} \langle c_q   Q^2\rangle$ & $0$\\
		\hhline{|=|=|=}
        Sum & $-3i\frac{e^2}{f_\pi}F^{\mu\nu}\widetilde{F}_{\mu\nu} \langle PQ^2 \rangle$ & $\frac{e}{f_\pi^3}\epsilon_{\mu\nu\rho\sigma}A^\mu \langle Q\partial^\nu \widetilde{P}\partial^\rho \widetilde{P}\partial^\sigma \widetilde{P}\rangle$ \\
        \hline
	\end{tabular}
	\caption{The contribution of the Wess-Zumino terms to the $PVV$ and $VPPP$ vertices, other terms are omitted.  
    $p.c.$ represents taking $\Sigma\leftrightarrow \Sigma^\dagger$ and $A_L\leftrightarrow A_R$, equivalent to replacing every pseudoscalar with minus itself. Note that the original terms are written in the language of differential forms, with wedges implicit. The sum of the terms, given in the last column, corresponds to \cref{eq:WZ_particular}.} 
	\label{tab:WZ_particular}
\end{table}	

The homogeneous Lagrangian is taken from Ref.~\cite{Fujiwara:1984mp}.
It consists of 6 terms, but only 3 are required to construct our model. 
They are given\footnote{Note that a factor of $g$ is missing from the definition $\cL_{\mathrm{hom},4}$ in Ref.~\cite{Fujiwara:1984mp}.} in \cref{tab:WZ_homogenous}, alongside their contributions to the to the $PVV$ and $VPPP$ vertices. 
We make use of the building blocks $\hat{\alpha}_{L,R}^\mu\equiv D^\mu(\xi_{L,R})\xi_{L,R}^\dagger-igV^\mu$ and $\hat{F}^{\mu\nu}_{L,R}\equiv eF^{\mu\nu}\xi_{L,R}Q\xi_{L,R}^\dagger$ therein, and follow the naming convention used by Ref.~\cite{Fujiwara:1984mp}.

\begin{table}[t]
	\centering
	\scriptsize
	\begin{tabular}{|c|c|c|}
		\hline
		Term & $PVV$ & $VPPP$ \\
		\hline
		$\cL_{\mathrm{\text{hom.}},1}=\frac{1}{2}  \epsilon_{\mu\nu\rho\sigma}\langle \hat{\alpha}_{L}^\mu\hat{\alpha}_{L}^\nu\hat{\alpha}_{L}^\rho\hat{\alpha}_{R}^\sigma\rangle-p.c.$ & $0$ & $\frac{2}{f_\pi^3}\epsilon_{\mu\nu\rho\sigma}\langle \left(gV^\mu -eQA^\mu\right)\widetilde{P}^\nu\widetilde{P}^\rho\widetilde{P}^\sigma\rangle$ \\
		\hline
		$\cL_{\mathrm{\text{hom.}},4}=\frac{1}{4} ig\epsilon_{\mu\nu\rho\sigma}\langle V^{\mu\nu}\hat{\alpha}_{L}^{\rho}\hat{\alpha}_{R}^{\sigma}\rangle-p.c.$ & $\frac{ig}{f_\pi}\langle \widetilde{P}\left(gV^{\mu\nu}\widetilde{V}_{\mu\nu}-\frac{1}{2} e\left\{V^{\mu\nu},Q\right\}\widetilde{F}_{\mu\nu}\right)\rangle$ & $\frac{2g}{f_\pi^3}\epsilon_{\mu\nu\rho\sigma}\langle V^\mu \widetilde{P}^\nu\widetilde{P}^\rho\widetilde{P}^\sigma\rangle$ \\
		\hline
		$\cL_{\mathrm{\text{hom.}},6}=\frac{1}{4} i\epsilon_{\mu\nu\rho\sigma}\langle \hat{F}_L^{\mu\nu}\hat{\alpha}_{L}^{\rho}\hat{\alpha}_{R}^{\sigma}\rangle-p.c.$ & $\frac{ie}{f_\pi}\widetilde{F}_{\mu\nu}\langle \widetilde{P}\left(\frac{1}{2} g\left\{V^{\mu\nu},Q\right\}-eQQF^{\mu\nu}\right)\rangle$ & $\frac{2e}{f_\pi^3}\epsilon_{\mu\nu\rho\sigma}\langle QA^\mu \widetilde{P}^\nu\widetilde{P}^\rho\widetilde{P}^\sigma\rangle$ \\
		\hline
	\end{tabular}
	\caption{The contribution of the Wess-Zumino homogeneous terms to the $PVV$ and $VPPP$ vertices. $p.c.$ represents taking $\Sigma\leftrightarrow \Sigma^\dagger$, $\xi_L\leftrightarrow \xi_R$ and $A_L\leftrightarrow A_R$}
	\label{tab:WZ_homogenous}
\end{table}	

Unlike the non-homogeneous terms, the coefficients of the homogeneous terms are free parameters since each term is individually invariant under the global symmetry.
The guiding principle in choosing these free coefficients is the VMD hypothesis.
By setting the coefficient of $\cL_4$ and $\cL_6$ to $c_4=c_6=-15C=\frac{iN_c}{16\pi^2}$, we achieve VMD for the pseudoscalar meson $PVV$ vertex, replacing the $P\gamma\gamma$ vertex from the non-homogeneous terms with a corresponding $PVV$ one.
The same can be done for the $VPPP$ interaction, replacing the interaction basis $\gamma PPP$ vertex by a $VPPP$ one.
However, Ref.~\cite{Fujiwara:1984mp} found that doing so leads to $\omega$ decay rates which are not in agreement with measurements.
Thus, they instead chose to eliminate the interaction basis $VPPP$ vertex in favor of a $\gamma PPP$ vertex, achieved by setting the coefficient of $\cL_1$ to $-\frac{iN_c}{16\pi^2}$.
Put together, the homogeneous Lagrangian contributes as
\begin{align}
	\cL_{\rm WZ}^{ \text{\tiny hom.}}
    =
    &-\frac{N_c}{16\pi^2f_\pi}\left(g^2 \langle \widetilde{P}V^{\mu\nu}\widetilde{V}_{\mu\nu}\rangle-e^2 F^{\mu\nu}\widetilde{F}_{\mu\nu}\langle \widetilde{P}QQ\rangle\right)\nonumber\\
    &\quad
    +\frac{ieN_c}{4\pi^2 f_\pi^3}\epsilon_{\mu\nu\rho\sigma}\langle QA^\mu \widetilde{P}^\nu\widetilde{P}^\rho\widetilde{P}^\sigma\rangle\eqd
\end{align}
Taken together with \cref{eq:WZ_particular} and $\cL_{a\gamma\gamma}$, we obtain \cref{eq:aVV_comp}. 

\subsection{Scalar resonances}
\label{AA:scalars}

We include the scalar resonances in our calculations, as they are found to have significant contributions to $a\rightarrow PPP$ processes~\cite{Aloni:2018vki}.
Following Ref.~\cite{Fariborz:1999gr,Black:1998wt}, we construct a nonet from the following scalar particles: 3 $a_0$ particles, 4 $\kappa$ particles, the $\sigma$ particle, and the $f_0$ particle.\footnote{These particles are now referred to as $a_0$, $K^*_0(700)$, $f_0(500)$ and $f_0(980)$, respectively. We use the older names to be consistent with~\cite{Fariborz:1999gr}.}
These are arranged into a  nonet in the following way: 
\begin{align}
	&N=\frac{1}{\sqrt{2}}\begin{pmatrix}
		\frac{-\sin(\theta_s)\sigma+\cos(\theta_s)f_0+a_0^0}{\sqrt{2}}&a_0^+&\kappa^+\\
		a_0^-&\frac{-\sin(\theta_s)\sigma+\cos(\theta_s)f_0-a_0^0}{\sqrt{2}}&\kappa^0\\
		\kappa^-&\bar{\kappa^0}&\cos(\theta_s)\sigma+\sin(\theta_s)f_0
	\end{pmatrix}\eqc
\end{align}
where $\theta_s=-20.33\degree$~\cite{Fariborz:1999gr}.

The Lagrangian contains the following terms relevant for the $SPP$ vertex:
\begin{align}
	\cL_{SPP}
	=&
	\sqrt{8}A\langle N  \partial_\mu \widetilde{P}\partial^\mu \widetilde{P}\rangle+
	\frac{\sqrt{8}(B-A)}{4}\langle N \rangle\langle \partial_\mu \widetilde{P}\partial^\mu \widetilde{P}\rangle+\nonumber\\
	&+\frac{\sqrt{8}(C-2A)}{4}\langle N \partial_\mu \widetilde{P}\rangle\langle \partial^\mu \widetilde{P}\rangle+
	\frac{\sqrt{8}(D+A)}{8}\langle N \rangle\langle \partial_\mu \widetilde{P}\rangle\langle \partial^\mu \widetilde{P}\rangle  
\end{align}
with $A=2.51\,\GeV^{-1}$, $B=-1.95\, \GeV^{-1}$, $C=7.16\,\GeV^{-1}$ and $D=-2.26\,\GeV^{-1}$ ~\cite{Fariborz:1999gr}.
This Lagrangian leads to an $SPP$ vertex of
\begin{align}
	\cL_{SPP}
	=
	\sum_s\sum_{p,p'}\frac{\gamma_{s,pp'}}{2}
	s \partial_\mu p \partial^\mu p'\eqc
\end{align}
with
\begin{align}
	\gamma_{s,pp'}=&\sqrt{8}A\langle T_s  \left\{T_p,T_{p'}\right\}\rangle+
	\sqrt{2}(B-A)\langle T_s \rangle\langle T_pT_{p'}\rangle+\nonumber\\
	&+\frac{\sqrt{2}(C-2A)}{2}\left(\langle T_s T_p\rangle\langle T_{p'}\rangle+\langle T_s T_{p'}\rangle\langle T_{p}\rangle\right)+
	\frac{\sqrt{2}(D+A)}{2}\langle T_s \rangle\langle T_p\rangle\langle T_{p'}\rangle\, .  
\end{align}
The $\gamma_{s,pp'}$ values are listed in \cref{tab:SPP_values}.
By completeness, $\gamma_{s,pp'}$ extends to the ALP, with $T_p$ replaced with $\widetilde{T}_a$, namely $\gamma_{s,pa} = \sum_{p'}\gamma_{s,pp'}\langle T_{p'}\widetilde{T}_a'\rangle$.

We note that Ref.~\cite{Fariborz:1999gr} uses an $\eta-\eta'$ mixing angle of $\theta_{\eta\eta'}=-17.7\degree$, slightly different from our value of $-19.5\degree$.
As some $SPP$ vertices are quite sensitive to this angle, we have used their mixing angle for these calculations only.

\begin{table}[hbt]
	\centering
	\scriptsize
	\begin{tabular}{|c|c||c|c||c|c|}
		\hline
		Vertex & Value & Vertex & Value & Vertex & Value \\
		\hhhline{|=|=||=|=||=|=|}
        $\sigma\pi\pi$&-10.3&$a_0\pi\eta$&+6.80&$\kappa K \eta$&+0.943\\
		\hline		
        $\sigma\eta\eta$&-8.22&$a_0\pi\eta'$&+7.80&$\kappa K \eta'$&+9.68\\
		\hline		
        $\sigma\eta\eta'$&-2.65&$a_0^0K^+K^-$&+3.55&$\kappa^\pm K^\mp \pi^0$&+3.55\\
		\hline		
        $\sigma\eta'\eta'$&+2.86&$a_0^0K^0\bar{K}^0$&-3.55&$\kappa^0 \bar{K}^0 \pi^0$& \multirow{2}{*}{-3.55}\\
		\hhline{|-|-|-|-|-|~}		
        $f_0\pi\pi$&-2.08&$\sigma KK$&-6.81&$\bar{\kappa}^0K^0\pi^0$& \\
		\hline		
        $f_0\eta\eta$&-3.44&$f_0 KK$&-7.15&$\kappa^+ \bar{K}^0\pi^-$& \multirow{4}{*}{+5.02}\\
		\hhline{|-|-|-|-|-|~}		
        $f_0\eta\eta'$&+9.01&$a_0^+K^-K^0$&\multirow{2}{*}{+5.02}&$\kappa^- K^0\pi^+$&    \\
		\hhline{|-|-|-|~|-|~}		
        $f_0\eta'\eta'$&-5.20&$a_0^-K^+\bar{K}^0$&&$\bar{\kappa}^0 K^+ \pi^-$&   \\
		\hhline{|-|-|-|-|-|~}		
        &&&&$\kappa^0 K^- \pi^+$&   \\
		\hline		
\end{tabular}
\caption{The $SPP$ couplings used in our calculations.
Note that they are normalized differently than in Ref~\cite{Fariborz:1999gr}.
}
\label{tab:SPP_values}
\end{table}

\subsection{Tensor resonances}
\label{AA:tensor}

Lastly, we consider the contribution of the tensor $f_2(1270)$ resonance to $a\rightarrow f_2 (\pi\pi) \eta^{(')}$ decays, as it is found to be a substantial contributor to similar decays of particles such as the $\eta_c$~\cite{BaBar:2021fkz}.

The Lagrangian of the tensor is given via\cite{Suzuki:1993zs,Han:1998sg,Katz:2005ir}
\begin{align}
	\cL_{f_2}&
    =
    -g_{f_2PP}\frac{f_\pi^2}{4}\langle
    \left(D_\mu \Sigma^\dagger D_\nu \Sigma 
    -\frac{1}{2}\eta_{\mu\nu}D_\alpha \Sigma^\dagger D^\alpha \Sigma 
    \right)T_{f_2}\rangle f_{2}^{\mu\nu}\eqc
\end{align}
where $T_{f_2}\equiv \frac{1}{2}\diag(1,1,0)$.
Expanding to leading order and utilizing the symmetry of $f_2$, we obtain the $TPP$ vertex
\begin{align}
	\cL_{f_2PP}&=-\frac{g_{f_2PP}}{2}\langle\left(\left\{D_\mu \widetilde{P}, D_\nu \widetilde{P} \right\} -\eta_{\mu\nu}D_\alpha \widetilde{P} D^\alpha \widetilde{P} \right)T_{f_2}\rangle f_{2}^{\mu\nu}\eqd
\end{align}
The value of $g_{f_2PP}$ is found by comparing the predicted $f_2\rightarrow \pi\pi$ rate 
\begin{align}
\Gamma_{f_2\rightarrow \pi\pi}=\frac{g_{f_2PP}^2}{640\pi}m_f^3\left(1-\frac{4m_\pi^2}{m_{f_2}^2}\right)^{2.5}
\end{align}
to the measured value, yielding $g_{f_2PP}=13.1\,\GeV^{-1} $.
The tensor propagator is taken to be
\begin{align}
	\langle f_2(k) f_2(-k) \rangle
    =&i \BW(k^2) B_{\mu\nu,\rho\sigma}(k) \, ,
\end{align}
where
\begin{align}
    B_{\mu\nu,\rho\sigma}(k)
    \equiv 
    &\left(\eta_{\mu\rho}-\frac{k_\mu k_\rho}{k^2}\right)\left(\eta_{\nu\sigma}-\frac{k_\nu k_\sigma}{k^2}\right)+\left(\eta_{\nu\rho}-\frac{k_\nu k_\rho}{k^2}\right)\left(\eta_{\mu\sigma}-\frac{k_\mu k_\sigma}{k^2}\right) \nonumber\\
    &-\frac{2}{3}\left(\eta_{\mu\nu}-\frac{k_\mu k_\nu}{k^2}\right)\left(\eta_{\rho\sigma}-\frac{k_\rho k_\sigma}{k^2}\right) \eqc
\end{align}
which agrees with the formalism used to study such decays empirically (note this is not the Unitary gauge propagator)~\cite{BaBar:2010wqe}.	

\section{\texorpdfstring{ALP $U(3)$ matrix and $\kappa$ Dependence}{ALP U(3) matrix and kappa Dependence}}
\label{A:Ta_kappa}

We seek to show that $T_a$ transforms under the field redefinition as $T_a\rightarrow T_a+\kappa$.
To do so, we first need to show that $h_{ai}\rightarrow h_{ai}+\langle \kappa T_i\rangle$, with the transformation of $T_a$ immediately following from completeness.
Since $m_{\eta\eta'}^2=0$ and $m_{\eta\pi}^2,m_{\eta'\pi}^2\propto\delta_I$, we shall work to first order in the meson mass mixing.
From \cref{eq:mix_kai,eq:mix_mai}, we see that the mixings transform as
\begin{align}
	&\delta K_{ai}
    =
    -\langle T_i\kappa \rangle\eqc\\
	&\delta m_{ai}^2
    =
    -2B_0\langle T_i \kappa M_q \rangle
    -\frac{m_0^2}{6}\langle T_i\rangle\langle \kappa\rangle\eqd
\end{align}
The second expression may be simplified by noting that the (flavor-neutral) meson mass matrix is
\begin{align}
		&m^2_{ij}=2B_0\langle T_i T_j M_q\rangle+\frac{m_0^2}{6}\langle T_i\rangle\langle T_j\rangle \quad\quad \eqc
        \label{eq:meson_mixing}\\
		&\sum_j m^2_{ij} \langle T_j\kappa\rangle = 2B_0\langle T_i \kappa M_q\rangle+\frac{m_0^2}{6}\langle T_i\rangle\langle \kappa\rangle=-\delta m_{ai}^2\eqc
	\end{align}
where $i,j=\pi,\eta,\eta'$ and we use completeness in the second equation.
Plugging these transformations into \cref{eq:mix_hai}, we get
\begin{align}
	\delta h_{ai}
    &=
    -\frac{\sum_j \left(m^2_{ij}\langle T_j\kappa\rangle\right)-m_a^2\langle T_i\kappa\rangle+\sum_{j\neq i}m^2_{ij}\frac{m^2_{j}\langle T_j\kappa\rangle-m_a^2\langle T_j\kappa\rangle}{m_a^2-m_j^2}}{m_a^2-m_i^2}+\cO(m_{ij}^2m_{jk}^2)\nonumber  \\
    &=\langle T_i\kappa\rangle+\cO(m_{ij}^2m_{jk}^2)\eqd
\end{align}
It is important to note that in \cref{eq:meson_mixing} we assumed that the meson masses and mixings are exactly equal to the tree-level values.
Deviation from that assumption will change the transformation properties of $T_a$, ultimately leading to basis-dependent processes.
As stated before, these unphysical dependencies are small.

\section{\texorpdfstring{ALP $U(3)$ matrix in the low and high energy limits}{ALP U(3) matrix in the low and high energy limits}}
\label{A:Ta_limits}

For $m_a\gg m_q$, \cref{eq:ta_cq} simplifies to
\begin{align}
	\widetilde{T}_a
    \approx   
    \frac{1}{m_a^2}\sum_i\left(m^2_{ai}-\sum_{j}m^2_{ij}K_{aj}
    \right)T_i\eqd
\end{align}
Utilizing \cref{eq:mix_mai,eq:mix_kai,eq:meson_mixing}, we get
\begin{align}
	\widetilde{T}_a\approx \frac{1}{m_a^2}
    \sum_i\Bigg[
    &\left( \frac{m_0^2}{6}c_g \langle T_i\rangle
    -2B_0\langle T_i\mang M_q\rangle \right) \nonumber\\
    &-\sum_{j}\left(2B_0\langle T_iT_j M_q\rangle+\frac{m_0^2}{6}\langle T_i \rangle\langle T_j \rangle\right)\langle T_j c_q\rangle
    \Bigg]T_i\eqd
\end{align}
Using completeness, we get
\begin{align}
    \label{eq:Ttileamid}
	\widetilde{T}_a
    \approx   
    \frac{1}{m_a^2}\left[\left(
    \frac{m_0^2}{6}c_g \idet
    -2B_0\mang M_q\right)
    -\left(2B_0c_q M_q+\frac{m_0^2}{6}\idet \langle c_q \rangle\right)
    \right]\eqd
\end{align}
Collecting terms, \cref{eq:Ttileamid} becomes
\begin{align}
	\widetilde{T}_a
    \approx   
    \frac{m_0^2}{6m_a^2}\left(c_g-\langle c_q\rangle\right)\idet-\frac{2B_0 M_q\left(\mang+c_q\right)}{m_a^2}\eqc
\end{align}
proving \cref{eq:Tatilhighmass}.

For $m_a\ll m_q$, $T_a$ becomes
\begin{align}
	T_a
    \approx   
    -\sum_i\frac{m^2_{ai}-\sum_{j\neq i}m^2_{ij}\frac{m^2_{aj}}{m_j^2}}{m_i^2}T_i\eqd
\end{align}
Since $T_a$ by itself is not invariant, it is worthwhile to work with an invariant object instead. Since $T_a$ does not depend on $c_q$ in this limit as all $c_q$ dependence comes from \cref{eq:mix_kai}, a convenient object to choose is $\bar{T}_a\equiv T_a-\mang$. 
Since it does not depend on $c_q$ and must be invariant, we immediately conclude it must be proportional to $\bar{c}_g\equiv c_g+\langle\mang\rangle$, which we explicitly verified.
\Cref{eq:ta_lowmass} then becomes evident by checking in a simple basis, such as the one where $\mang=0$.

\section{\texorpdfstring{$a\rightarrow VP$}{a -> VP}}
\label{A:decay_VP}

In this appendix, we study the decay of the ALP into a vector and a pseudoscalar.
These processes contribute to the decay of an ALP into 3 pseudoscalars via $a\rightarrow V(PP)P$.
Therefore, to avoid double counting, they are not included in the ALP total decay rate, and are merely given for here completeness.
Such decays come from the vertex detailed in \cref{eq:VPP}.
The rate is found to be
\begin{align}
	\Gamma_{a\rightarrow VP}
	=
	\frac{\alpha_g f_\pi^2 m_a^3}{4 f_a^2 m_V^2}&\cF_{VPP}^2
	\left[\langle\left(T_a+c_q\right)
	\left[T_P,T_V\right]\rangle\right]^2\nonumber\\
	&\times\left[\left(1-\frac{\left(m_V+m_P\right)^2}{m_a^2}\right)
	\left(1-\frac{\left(m_V-m_P\right)^2}{m_a^2}\right)\right]^{3/2}\eqc
\end{align}
with $T_P,T_V$ being respectively the pseudoscalar and vector matrices, and $m_P$, $m_V$ being their masses.	
Due to the commutator, decays into flavor-neutral mesons or a photon are disallowed.
Additionally, all $a\rightarrow VP$ decays violate $U(3)_V$, and thus vanish for the case that the ALP couples solely to gluons.

\section{The \texorpdfstring{$\kappa$}{kappa} Breit Wigner}
\label{A:kappa_BW}

%
\begin{figure}[t]
  \centering
    \includegraphics[width=0.99\textwidth]
    {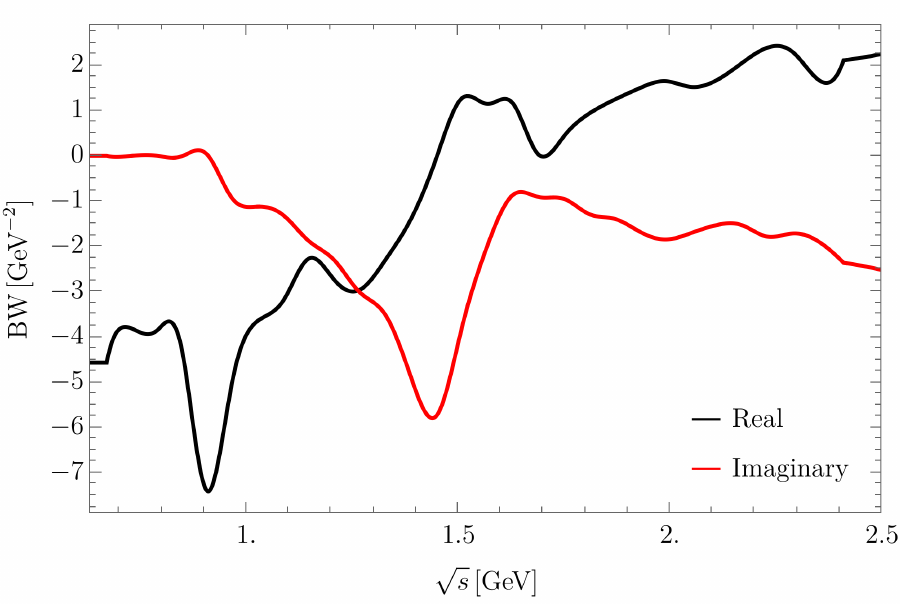}
      \caption{The real and imaginary parts of the $\kappa$ propagator.}
    \label{fig:kappa_BW}
\end{figure}

It is known that $\kappa$ mediated processes receive additional contributions from higher order resonances, most notably $K^*(1430)$, with the overall propagator not being well described as a simple sum of Breit Wigner terms.
As such, we use the amplitude measured by BaBar using $\eta_c\to KK\pi$ data~\cite{BaBar:2015kii}.
The decay of the ALP (and $\eta_c$) into $KK\pi$ is facilitated by two $\kappa$ diagrams and an $a_0$ diagram, with the former being of interest to us.
The isospin symmetry implies that the two $\kappa$ diagrams have the same couplings and the same BW distributions.

The individual amplitudes take the form
\begin{align}
    &\cM
    =
    \frac{\gamma_{\bar{\kappa},K\pi}\gamma_{\kappa,Ka}}{4} (m_a^2-s+m_K^2)(s-m_\pi^2-m_K^2) \widehat{\BW}(s)\cF_{SPP}(m_a)\eqc
\end{align}
where $\widehat{\BW}(s)\equiv \BW(s)\cF_{SPP}(\sqrt{s})$, $s$ is the squared invariant mass of the meson pair corresponding to the $\kappa$ and we have explicitly performed the momentum contractions of \cref{eq:4P_s_mediated}.

To estimate the $\eta_c\rightarrow KK\pi$ amplitude, we replace $m_a$ with $m_{\eta_c}$, and use $\widetilde{T_a}=\frac{\alpha_s(m_{\eta_c})}{\sqrt{6}}\idet$ when calculating $\gamma_{\kappa,Ka}$.
As a result, $\cF_{SPP}(m_a)$ and $\gamma_{\kappa,Ka}$ are simply constants that do not depend on the kinematics, and we find
\begin{align}
    &\widehat{\BW}(s)
    \propto
    \frac{\cM_{\eta_c\rightarrow \kappa(K\pi)K}}{(m_{\eta_c}^2-s+m_K^2)(s-m_\pi^2-m_K^2)}\eqd
\end{align}
The measured amplitude $\cM_{\eta_c\rightarrow \kappa(K\pi)K}$ is taken from Ref.~\cite{BaBar:2015kii}, and we averaged the measurements coming from $\eta_c\rightarrow K_s^0 K^\pm \pi^\mp$ and $\eta_c\rightarrow K^\pm K^\mp \pi^0$.
Ref.~\cite{BaBar:2015kii} set their phase to a value of $\pi/2$ at 1450\,MeV, corresponding to the BW pole of $K^*(1430)$, which we follow (up to a sign coming from the sign convention of the BW).

Lastly, we note that it is not well motivated to use the $\kappa$ couplings of Ref.~\cite{Fariborz:1999gr} for this complex distribution comprised of multiple resonances.
Therefore, we  rescale the vertices of the $K\pi$ S-wave by a constant factor such that our prediction matches the measured $\eta_c\rightarrow KK\pi$ branching fraction, for which the $K\pi$ S-wave is the main contribution~\cite{BaBar:2015kii}.
The resulting BW function, multiplied by the aforementioned rescaling factor, is displayed in~\cref{fig:kappa_BW}.

\bibliographystyle{JHEP.bst}
\bibliography{bibli.bib}

@article{Chetyrkin:2000yt,
    author = "Chetyrkin, K. G. and Kuhn, Johann H. and Steinhauser, M.",
    title = "{RunDec: A Mathematica package for running and decoupling of the strong coupling and quark masses}",
    eprint = "hep-ph/0004189",
    archivePrefix = "arXiv",
    reportNumber = "DESY-00-034, TTP-00-05",
    doi = "10.1016/S0010-4655(00)00155-7",
    journal = "Comput. Phys. Commun.",
    volume = "133",
    pages = "43--65",
    year = "2000"
}

@article{Bai:2024lpq,
    author = "Bai, Yang and Chen, Ting-Kuo and Liu, Jia and Ma, Xiaolin",
    title = "{Wess-Zumino-Witten Interactions of Axions}",
    eprint = "2406.11948",
    archivePrefix = "arXiv",
    primaryClass = "hep-ph",
    doi = "10.1103/PhysRevLett.134.081803",
    journal = "Phys. Rev. Lett.",
    volume = "134",
    number = "8",
    pages = "081803",
    year = "2025"
}

@article{BaBar:2010wqe,
    author = "del Amo Sanchez, P. and others",
    collaboration = "BaBar",
    title = "{Dalitz plot analysis of $D_s^+ \to K^+ K^- \pi^+$}",
    eprint = "1011.4190",
    archivePrefix = "arXiv",
    primaryClass = "hep-ex",
    reportNumber = "BABAR-PUB-10-016, SLAC-PUB-14266",
    doi = "10.1103/PhysRevD.83.052001",
    journal = "Phys. Rev. D",
    volume = "83",
    pages = "052001",
    year = "2011"
}

@article{Blinov:2021say,
    author = "Blinov, Nikita and Kowalczyk, Elizabeth and Wynne, Margaret",
    title = "{Axion-like particle searches at DarkQuest}",
    eprint = "2112.09814",
    archivePrefix = "arXiv",
    primaryClass = "hep-ph",
    reportNumber = "FERMILAB-PUB-21-749-V",
    doi = "10.1007/JHEP02(2022)036",
    journal = "JHEP",
    volume = "02",
    pages = "036",
    year = "2022"
}

@article{Gasser:1984gg,
	author = "Gasser, J. and Leutwyler, H.",
	title = "{Chiral Perturbation Theory: Expansions in the Mass of the Strange Quark}",
	reportNumber = "CERN-TH-3798",
	doi = "10.1016/0550-3213(85)90492-4",
	journal = "Nucl. Phys. B",
	volume = "250",
	pages = "465--516",
	year = "1985"
}

@article{Bai:2025fvl,
    author = "Bai, Yang and Chen, Ting-Kuo and Liu, Jia and Ma, Xiaolin",
    title = "{Wess-Zumino-Witten Interactions of Axions: Three-Flavor}",
    eprint = "2505.24822",
    archivePrefix = "arXiv",
    primaryClass = "hep-ph",
    month = "5",
    year = "2025"
}

@article{Bando:1984ej,
	author = "Bando, M. and Kugo, T. and Uehara, S. and Yamawaki, K. and Yanagida, T.",
	title = "{Is rho Meson a Dynamical Gauge Boson of Hidden Local Symmetry?}",
	reportNumber = "RRK 84-22",
	doi = "10.1103/PhysRevLett.54.1215",
	journal = "Phys. Rev. Lett.",
	volume = "54",
	pages = "1215",
	year = "1985"
}

@article{Fujiwara:1984mp,
	author = "Fujiwara, Takanori and Kugo, Taichiro and Terao, Haruhiko and Uehara, Shozo and Yamawaki, Koichi",
	title = "{Nonabelian Anomaly and Vector Mesons as Dynamical Gauge Bosons of Hidden Local Symmetries}",
	reportNumber = "KUNS-764",
	doi = "10.1143/PTP.73.926",
	journal = "Prog. Theor. Phys.",
	volume = "73",
	pages = "926",
	year = "1985"
}

@article{Callan:1969sn,
    author = "Callan, Jr., Curtis G. and Coleman, Sidney R. and Wess, J. and Zumino, Bruno",
    title = "{Structure of phenomenological Lagrangians. 2.}",
    doi = "10.1103/PhysRev.177.2247",
    journal = "Phys. Rev.",
    volume = "177",
    pages = "2247--2250",
    year = "1969"
}

@article{Bauer:2021wjo,
	author = "Bauer, Martin and Neubert, Matthias and Renner, Sophie and Schnubel, Marvin and Thamm, Andrea",
	title = "{Consistent Treatment of Axions in the Weak Chiral Lagrangian}",
	eprint = "2102.13112",
	archivePrefix = "arXiv",
	primaryClass = "hep-ph",
	reportNumber = "IPPP/20-82, MITP/21-007, ZU-TH-01/21",
	doi = "10.1103/PhysRevLett.127.081803",
	journal = "Phys. Rev. Lett.",
	volume = "127",
	number = "8",
	pages = "081803",
	year = "2021"
}

@article{Georgi:1986df,
	author = "Georgi, Howard and Kaplan, David B. and Randall, Lisa",
	title = "{Manifesting the Invisible Axion at Low-energies}",
	reportNumber = "HUTP-86/A004",
	doi = "10.1016/0370-2693(86)90688-X",
	journal = "Phys. Lett. B",
	volume = "169",
	pages = "73--78",
	year = "1986"
}

@article{Witten:1983tw,
	author = "Witten, Edward",
	title = "{Global Aspects of Current Algebra}",
	reportNumber = "PRINT-83-0262 (PRINCETON)",
	doi = "10.1016/0550-3213(83)90063-9",
	journal = "Nucl. Phys. B",
	volume = "223",
	pages = "422--432",
	year = "1983"
}

@article{Kaymakcalan:1983qq,
	author = "Kaymakcalan, O. and Rajeev, S. and Schechter, J.",
	title = "{Nonabelian Anomaly and Vector Meson Decays}",
	reportNumber = "SU-4222-278, COO-3533-278",
	doi = "10.1103/PhysRevD.30.594",
	journal = "Phys. Rev. D",
	volume = "30",
	pages = "594",
	year = "1984"
}

@article{Aloni:2018vki,
	author = "Aloni, Daniel and Soreq, Yotam and Williams, Mike",
	title = "{Coupling QCD-Scale Axionlike Particles to Gluons}",
	eprint = "1811.03474",
	archivePrefix = "arXiv",
	primaryClass = "hep-ph",
	reportNumber = "CERN-TH-2018-237, MIT-CTP/5080, MIT-CTP-5080",
	doi = "10.1103/PhysRevLett.123.031803",
	journal = "Phys. Rev. Lett.",
	volume = "123",
	number = "3",
	pages = "031803",
	year = "2019"
}

@article{Cheng:2021kjg,
	author = "Cheng, Hsin-Chia and Li, Lingfeng and Salvioni, Ennio",
	title = "{A theory of dark pions}",
	eprint = "2110.10691",
	archivePrefix = "arXiv",
	primaryClass = "hep-ph",
	reportNumber = "CERN-TH-2021-150",
	doi = "10.1007/JHEP01(2022)122",
	journal = "JHEP",
	volume = "01",
	pages = "122",
	year = "2022"
}

@article{Bauer:2021mvw,
	author = "Bauer, Martin and Neubert, Matthias and Renner, Sophie and Schnubel, Marvin and Thamm, Andrea",
	title = "{Flavor probes of axion-like particles}",
	eprint = "2110.10698",
	archivePrefix = "arXiv",
	primaryClass = "hep-ph",
	reportNumber = "MITP/21-025, CERN-TH-2021-148, IPPP/21/37",
	doi = "10.1007/JHEP09(2022)056",
	journal = "JHEP",
	volume = "09",
	pages = "056",
	year = "2022"
}

@article{ParticleDataGroup:2022pth,
	author = "Workman, R. L. and others",
	collaboration = "Particle Data Group",
	title = "{Review of Particle Physics}",
	doi = "10.1093/ptep/ptac097",
	journal = "PTEP",
	volume = "2022",
	pages = "083C01",
	year = "2022"
}

@article{Peccei:1977hh,
	author = "Peccei, R. D. and Quinn, Helen R.",
	title = "{CP Conservation in the Presence of Instantons}",
	reportNumber = "ITP-568-STANFORD",
	doi = "10.1103/PhysRevLett.38.1440",
	journal = "Phys. Rev. Lett.",
	volume = "38",
	pages = "1440--1443",
	year = "1977"
}

@article{Peccei:1977ur,
	author = "Peccei, R. D. and Quinn, Helen R.",
	title = "{Constraints Imposed by CP Conservation in the Presence of Instantons}",
	reportNumber = "ITP-572-STANFORD",
	doi = "10.1103/PhysRevD.16.1791",
	journal = "Phys. Rev. D",
	volume = "16",
	pages = "1791--1797",
	year = "1977"
}

@article{Weinberg:1977ma,
	author = "Weinberg, Steven",
	title = "{A New Light Boson?}",
	reportNumber = "HUTP-77/A074",
	doi = "10.1103/PhysRevLett.40.223",
	journal = "Phys. Rev. Lett.",
	volume = "40",
	pages = "223--226",
	year = "1978"
}

@article{Wilczek:1977pj,
	author = "Wilczek, Frank",
	title = "{Problem of Strong  $P$  and  $T$  Invariance in the Presence of Instantons}",
	reportNumber = "Print-77-0939 (COLUMBIA)",
	doi = "10.1103/PhysRevLett.40.279",
	journal = "Phys. Rev. Lett.",
	volume = "40",
	pages = "279--282",
	year = "1978"
}

@article{Nomura:2008ru,
	author = "Nomura, Yasunori and Thaler, Jesse",
	title = "{Dark Matter through the Axion Portal}",
	eprint = "0810.5397",
	archivePrefix = "arXiv",
	primaryClass = "hep-ph",
	doi = "10.1103/PhysRevD.79.075008",
	journal = "Phys. Rev. D",
	volume = "79",
	pages = "075008",
	year = "2009"
}

@article{Freytsis:2010ne,
	author = "Freytsis, Marat and Ligeti, Zoltan",
	title = "{On dark matter models with uniquely spin-dependent detection possibilities}",
	eprint = "1012.5317",
	archivePrefix = "arXiv",
	primaryClass = "hep-ph",
	doi = "10.1103/PhysRevD.83.115009",
	journal = "Phys. Rev. D",
	volume = "83",
	pages = "115009",
	year = "2011"
}

@article{Dolan:2014ska,
	author = "Dolan, Matthew J. and Kahlhoefer, Felix and McCabe, Christopher and Schmidt-Hoberg, Kai",
	title = "{A taste of dark matter: Flavour constraints on pseudoscalar mediators}",
	eprint = "1412.5174",
	archivePrefix = "arXiv",
	primaryClass = "hep-ph",
	reportNumber = "DESY-14-238, SLAC-PUB-16179",
	doi = "10.1007/JHEP03(2015)171",
	journal = "JHEP",
	volume = "03",
	pages = "171",
	year = "2015",
	note = "[Erratum: JHEP 07, 103 (2015)]"
}

@article{Hochberg:2018rjs,
	author = "Hochberg, Yonit and Kuflik, Eric and Mcgehee, Robert and Murayama, Hitoshi and Schutz, Katelin",
	title = "{Strongly interacting massive particles through the axion portal}",
	eprint = "1806.10139",
	archivePrefix = "arXiv",
	primaryClass = "hep-ph",
	reportNumber = "DESY-18-101, IPMU18-0114",
	doi = "10.1103/PhysRevD.98.115031",
	journal = "Phys. Rev. D",
	volume = "98",
	number = "11",
	pages = "115031",
	year = "2018"
}

@article{Preskill:1982cy,
	author = "Preskill, John and Wise, Mark B. and Wilczek, Frank",
	editor = "Srednicki, M. A.",
	title = "{Cosmology of the Invisible Axion}",
	reportNumber = "HUTP-82-A048, NSF-ITP-82-103",
	doi = "10.1016/0370-2693(83)90637-8",
	journal = "Phys. Lett. B",
	volume = "120",
	pages = "127--132",
	year = "1983"
}

@article{Abbott:1982af,
	author = "Abbott, L. F. and Sikivie, P.",
	editor = "Srednicki, M. A.",
	title = "{A Cosmological Bound on the Invisible Axion}",
	reportNumber = "PRINT-82-0695 (BRANDEIS)",
	doi = "10.1016/0370-2693(83)90638-X",
	journal = "Phys. Lett. B",
	volume = "120",
	pages = "133--136",
	year = "1983"
}

@article{Dine:1982ah,
	author = "Dine, Michael and Fischler, Willy",
	editor = "Srednicki, M. A.",
	title = "{The Not So Harmless Axion}",
	reportNumber = "UPR-0201T",
	doi = "10.1016/0370-2693(83)90639-1",
	journal = "Phys. Lett. B",
	volume = "120",
	pages = "137--141",
	year = "1983"
}

@article{Graham:2015cka,
	author = "Graham, Peter W. and Kaplan, David E. and Rajendran, Surjeet",
	title = "{Cosmological Relaxation of the Electroweak Scale}",
	eprint = "1504.07551",
	archivePrefix = "arXiv",
	primaryClass = "hep-ph",
	doi = "10.1103/PhysRevLett.115.221801",
	journal = "Phys. Rev. Lett.",
	volume = "115",
	number = "22",
	pages = "221801",
	year = "2015"
}

@article{Lepage:1980fj,
	author = "Lepage, G. Peter and Brodsky, Stanley J.",
	title = "{Exclusive Processes in Perturbative Quantum Chromodynamics}",
	reportNumber = "SLAC-PUB-2478",
	doi = "10.1103/PhysRevD.22.2157",
	journal = "Phys. Rev. D",
	volume = "22",
	pages = "2157",
	year = "1980"
}

@article{GrillidiCortona:2015jxo,
    author = "Grilli di Cortona, Giovanni and Hardy, Edward and Pardo Vega, Javier and Villadoro, Giovanni",
    title = "{The QCD axion, precisely}",
    eprint = "1511.02867",
    archivePrefix = "arXiv",
    primaryClass = "hep-ph",
    doi = "10.1007/JHEP01(2016)034",
    journal = "JHEP",
    volume = "01",
    pages = "034",
    year = "2016"
}

@article{Fariborz:1999gr,
	author = "Fariborz, Amir H. and Schechter, Joseph",
	title = "{Eta-prime ---\ensuremath{>} eta pi pi decay as a probe of a possible lowest lying scalar nonet}",
	eprint = "hep-ph/9902238",
	archivePrefix = "arXiv",
	reportNumber = "SU-4240-693",
	doi = "10.1103/PhysRevD.60.034002",
	journal = "Phys. Rev. D",
	volume = "60",
	pages = "034002",
	year = "1999"
}

@article{BaBar:2012bdw,
    author = "Lees, J. P. and others",
    collaboration = "BaBar",
    title = "{Precise Measurement of the $e^+ e^- \to \pi^+\pi^- (\gamma)$ Cross Section with the Initial-State Radiation Method at BABAR}",
    eprint = "1205.2228",
    archivePrefix = "arXiv",
    primaryClass = "hep-ex",
    reportNumber = "BABAR-PUB-12-003",
    doi = "10.1103/PhysRevD.86.032013",
    journal = "Phys. Rev. D",
    volume = "86",
    pages = "032013",
    year = "2012"
}

@article{BaBar:2015kii,
    author = "Lees, J. P. and others",
    collaboration = "BaBar",
    title = "{Measurement of the I=1/2 $K \pi$ $\mathcal{S}$-wave amplitude from Dalitz plot analyses of $\eta_c \to K \bar K \pi$ in two-photon interactions}",
    eprint = "1511.02310",
    archivePrefix = "arXiv",
    primaryClass = "hep-ex",
    reportNumber = "BABAR-PUB-15-008, SLAC-PUB-16422",
    doi = "10.1103/PhysRevD.93.012005",
    journal = "Phys. Rev. D",
    volume = "93",
    pages = "012005",
    year = "2016"
}

@article{Bauer:2020jbp,
    author = "Bauer, Martin and Neubert, Matthias and Renner, Sophie and Schnubel, Marvin and Thamm, Andrea",
    title = "{The Low-Energy Effective Theory of Axions and ALPs}",
    eprint = "2012.12272",
    archivePrefix = "arXiv",
    primaryClass = "hep-ph",
    reportNumber = "IPPP/20/69, MITP/20-070 SISSA 30/2020/FISI, ZH-TH-47/20",
    doi = "10.1007/JHEP04(2021)063",
    journal = "JHEP",
    volume = "04",
    pages = "063",
    year = "2021"
}

@article{MartinCamalich:2020dfe,
    author = "Martin Camalich, Jorge and Pospelov, Maxim and Vuong, Pham Ngoc Hoa and Ziegler, Robert and Zupan, Jure",
    title = "{Quark Flavor Phenomenology of the QCD Axion}",
    eprint = "2002.04623",
    archivePrefix = "arXiv",
    primaryClass = "hep-ph",
    doi = "10.1103/PhysRevD.102.015023",
    journal = "Phys. Rev. D",
    volume = "102",
    number = "1",
    pages = "015023",
    year = "2020"
}

@article{Spira:1995rr,
    author = "Spira, M. and Djouadi, A. and Graudenz, D. and Zerwas, P. M.",
    title = "{Higgs boson production at the LHC}",
    eprint = "hep-ph/9504378",
    archivePrefix = "arXiv",
    reportNumber = "DESY-94-123, UDEM-GPP-TH-95-16, CERN-TH-95-30, CERN-TH-95-030",
    doi = "10.1016/0550-3213(95)00379-7",
    journal = "Nucl. Phys. B",
    volume = "453",
    pages = "17--82",
    year = "1995"
}

@article{Djouadi:2005gj,
    author = "Djouadi, Abdelhak",
    title = "{The Anatomy of electro-weak symmetry breaking. II. The Higgs bosons in the minimal supersymmetric model}",
    eprint = "hep-ph/0503173",
    archivePrefix = "arXiv",
    reportNumber = "LPT-ORSAY-05-18",
    doi = "10.1016/j.physrep.2007.10.005",
    journal = "Phys. Rept.",
    volume = "459",
    pages = "1--241",
    year = "2008"
}

@article{Coleman:1969sm,
    author = "Coleman, Sidney R. and Wess, J. and Zumino, Bruno",
    title = "{Structure of phenomenological Lagrangians. 1.}",
    doi = "10.1103/PhysRev.177.2239",
    journal = "Phys. Rev.",
    volume = "177",
    pages = "2239--2247",
    year = "1969"
}

@article{Sakurai:1960ju,
    author = "Sakurai, J. J.",
    title = "{Theory of strong interactions}",
    doi = "10.1016/0003-4916(60)90126-3",
    journal = "Annals Phys.",
    volume = "11",
    pages = "1--48",
    year = "1960"
}

@article{Fitzpatrick:2023xks,
    author = "Fitzpatrick, Patrick J. and Hochberg, Yonit and Kuflik, Eric and Ovadia, Rotem and Soreq, Yotam",
    title = "{Dark matter through the axion-gluon portal}",
    eprint = "2306.03128",
    archivePrefix = "arXiv",
    primaryClass = "hep-ph",
    doi = "10.1103/PhysRevD.108.075003",
    journal = "Phys. Rev. D",
    volume = "108",
    number = "7",
    pages = "075003",
    year = "2023"
}

@article{Agrawal:2017eqm,
    author = "Agrawal, Prateek and Marques-Tavares, Gustavo and Xue, Wei",
    title = "{Opening up the QCD axion window}",
    eprint = "1708.05008",
    archivePrefix = "arXiv",
    primaryClass = "hep-ph",
    doi = "10.1007/JHEP03(2018)049",
    journal = "JHEP",
    volume = "03",
    pages = "049",
    year = "2018"
}

@article{Agrawal:2017ksf,
    author = "Agrawal, Prateek and Howe, Kiel",
    title = "{Factoring the Strong CP Problem}",
    eprint = "1710.04213",
    archivePrefix = "arXiv",
    primaryClass = "hep-ph",
    reportNumber = "FERMILAB-PUB-17-500-T",
    doi = "10.1007/JHEP12(2018)029",
    journal = "JHEP",
    volume = "12",
    pages = "029",
    year = "2018"
}

@article{Gaillard:2018xgk,
    author = "Gaillard, M. K. and Gavela, M. B. and Houtz, R. and Quilez, P. and Del Rey, R.",
    title = "{Color unified dynamical axion}",
    eprint = "1805.06465",
    archivePrefix = "arXiv",
    primaryClass = "hep-ph",
    reportNumber = "IFT-UAM/CSIC-18-050, FTUAM-18-13, IFT-UAM-CSIC-18-050",
    doi = "10.1140/epjc/s10052-018-6396-6",
    journal = "Eur. Phys. J. C",
    volume = "78",
    number = "11",
    pages = "972",
    year = "2018"
}

@article{Gherghetta:2020keg,
    author = "Gherghetta, Tony and Khoze, Valentin V. and Pomarol, Alex and Shirman, Yuri",
    title = "{The Axion Mass from 5D Small Instantons}",
    eprint = "2001.05610",
    archivePrefix = "arXiv",
    primaryClass = "hep-ph",
    reportNumber = "UMN-TH-3909/20, IPPP-19-95",
    doi = "10.1007/JHEP03(2020)063",
    journal = "JHEP",
    volume = "03",
    pages = "063",
    year = "2020"
}

@article{Gupta:2020vxb,
    author = "Gupta, R. S. and Khoze, V. V. and Spannowsky, M.",
    title = "{Small instantons and the strong CP problem in composite Higgs models}",
    eprint = "2012.00017",
    archivePrefix = "arXiv",
    primaryClass = "hep-ph",
    doi = "10.1103/PhysRevD.104.075011",
    journal = "Phys. Rev. D",
    volume = "104",
    number = "7",
    pages = "075011",
    year = "2021"
}

@article{Gherghetta:2020ofz,
    author = "Gherghetta, Tony and Nguyen, Minh D.",
    title = "{A Composite Higgs with a Heavy Composite Axion}",
    eprint = "2007.10875",
    archivePrefix = "arXiv",
    primaryClass = "hep-ph",
    reportNumber = "UMN--TH--3923/20",
    doi = "10.1007/JHEP12(2020)094",
    journal = "JHEP",
    volume = "12",
    pages = "094",
    year = "2020"
}

@article{Valenti:2022tsc,
    author = "Valenti, Alessandro and Vecchi, Luca and Xu, Ling-Xiao",
    title = "{Grand Color axion}",
    eprint = "2206.04077",
    archivePrefix = "arXiv",
    primaryClass = "hep-ph",
    doi = "10.1007/JHEP10(2022)025",
    journal = "JHEP",
    volume = "10",
    pages = "025",
    year = "2022"
}

@article{Ovchynnikov:2025gpx,
    author = "Ovchynnikov, Maksym and Zaporozhchenko, Andrii",
    title = "{ALPs coupled to gluons in the GeV mass range -- data-driven and consistent}",
    eprint = "2501.04525",
    archivePrefix = "arXiv",
    primaryClass = "hep-ph",
    reportNumber = "CERN-TH-2025-006",
    month = "1",
    year = "2025"
}

@article{BaBar:2021fkz,
    author = "Lees, J. P. and others",
    collaboration = "BaBar",
    title = "{Light meson spectroscopy from Dalitz plot analyses of $\eta_c$ decays to $\eta' K^+ K^-$, $\eta' \pi^+ \pi^-$, and $\eta \pi^+ \pi^-$ produced in two-photon interactions}",
    eprint = "2106.05157",
    archivePrefix = "arXiv",
    primaryClass = "hep-ex",
    reportNumber = "BaBar-PUB-21/001,SLAC-PUB-17606",
    doi = "10.1103/PhysRevD.104.072002",
    journal = "Phys. Rev. D",
    volume = "104",
    number = "7",
    pages = "072002",
    year = "2021"
}

@article{Marsh:2015xka,
    author = "Marsh, David J. E.",
    title = "{Axion Cosmology}",
    eprint = "1510.07633",
    archivePrefix = "arXiv",
    primaryClass = "astro-ph.CO",
    reportNumber = "KCL-PH-TH-2015-50",
    doi = "10.1016/j.physrep.2016.06.005",
    journal = "Phys. Rept.",
    volume = "643",
    pages = "1--79",
    year = "2016"
}

@article{Graham:2015ouw,
    author = "Graham, Peter W. and Irastorza, Igor G. and Lamoreaux, Steven K. and Lindner, Axel and van Bibber, Karl A.",
    title = "{Experimental Searches for the Axion and Axion-Like Particles}",
    eprint = "1602.00039",
    archivePrefix = "arXiv",
    primaryClass = "hep-ex",
    doi = "10.1146/annurev-nucl-102014-022120",
    journal = "Ann. Rev. Nucl. Part. Sci.",
    volume = "65",
    pages = "485--514",
    year = "2015"
}

@article{Hook:2018dlk,
    author = "Hook, Anson",
    title = "{TASI Lectures on the Strong CP Problem and Axions}",
    eprint = "1812.02669",
    archivePrefix = "arXiv",
    primaryClass = "hep-ph",
    doi = "10.22323/1.333.0004",
    journal = "PoS",
    volume = "TASI2018",
    pages = "004",
    year = "2019"
}

@article{Irastorza:2018dyq,
    author = "Irastorza, Igor G. and Redondo, Javier",
    title = "{New experimental approaches in the search for axion-like particles}",
    eprint = "1801.08127",
    archivePrefix = "arXiv",
    primaryClass = "hep-ph",
    doi = "10.1016/j.ppnp.2018.05.003",
    journal = "Prog. Part. Nucl. Phys.",
    volume = "102",
    pages = "89--159",
    year = "2018"
}

@article{Agrawal:2021dbo,
    author = "Agrawal, Prateek and others",
    title = "{Feebly-interacting particles: FIPs 2020 workshop report}",
    eprint = "2102.12143",
    archivePrefix = "arXiv",
    primaryClass = "hep-ph",
    doi = "10.1140/epjc/s10052-021-09703-7",
    journal = "Eur. Phys. J. C",
    volume = "81",
    number = "11",
    pages = "1015",
    year = "2021"
}

@article{Dolan:2017osp,
    author = "Dolan, Matthew J. and Ferber, Torben and Hearty, Christopher and Kahlhoefer, Felix and Schmidt-Hoberg, Kai",
    title = "{Revised constraints and Belle II sensitivity for visible and invisible axion-like particles}",
    eprint = "1709.00009",
    archivePrefix = "arXiv",
    primaryClass = "hep-ph",
    reportNumber = "DESY-17-127",
    doi = "10.1007/JHEP12(2017)094",
    journal = "JHEP",
    volume = "12",
    pages = "094",
    year = "2017",
    note = "[Erratum: JHEP 03, 190 (2021)]"
}

@article{Alves:2017avw,
    author = "Alves, Daniele S. M. and Weiner, Neal",
    title = "{A viable QCD axion in the MeV mass range}",
    eprint = "1710.03764",
    archivePrefix = "arXiv",
    primaryClass = "hep-ph",
    reportNumber = "LA-UR-17-29295",
    doi = "10.1007/JHEP07(2018)092",
    journal = "JHEP",
    volume = "07",
    pages = "092",
    year = "2018"
}

@article{Marciano:2016yhf,
    author = "Marciano, W. J. and Masiero, A. and Paradisi, P. and Passera, M.",
    title = "{Contributions of axionlike particles to lepton dipole moments}",
    eprint = "1607.01022",
    archivePrefix = "arXiv",
    primaryClass = "hep-ph",
    doi = "10.1103/PhysRevD.94.115033",
    journal = "Phys. Rev. D",
    volume = "94",
    number = "11",
    pages = "115033",
    year = "2016"
}

@article{Jaeckel:2015jla,
    author = "Jaeckel, Joerg and Spannowsky, Michael",
    title = "{Probing MeV to 90 GeV axion-like particles with LEP and LHC}",
    eprint = "1509.00476",
    archivePrefix = "arXiv",
    primaryClass = "hep-ph",
    doi = "10.1016/j.physletb.2015.12.037",
    journal = "Phys. Lett. B",
    volume = "753",
    pages = "482--487",
    year = "2016"
}

@article{Dobrich:2015jyk,
    author = {D\"obrich, Babette and Jaeckel, Joerg and Kahlhoefer, Felix and Ringwald, Andreas and Schmidt-Hoberg, Kai},
    title = "{ALPtraum: ALP production in proton beam dump experiments}",
    eprint = "1512.03069",
    archivePrefix = "arXiv",
    primaryClass = "hep-ph",
    reportNumber = "CERN-PH-TH-2015-293, DESY-15-237",
    doi = "10.1007/JHEP02(2016)018",
    journal = "JHEP",
    volume = "02",
    pages = "018",
    year = "2016"
}

@article{Izaguirre:2016dfi,
    author = "Izaguirre, Eder and Lin, Tongyan and Shuve, Brian",
    title = "{Searching for Axionlike Particles in Flavor-Changing Neutral Current Processes}",
    eprint = "1611.09355",
    archivePrefix = "arXiv",
    primaryClass = "hep-ph",
    reportNumber = "SLAC-PUB-16876",
    doi = "10.1103/PhysRevLett.118.111802",
    journal = "Phys. Rev. Lett.",
    volume = "118",
    number = "11",
    pages = "111802",
    year = "2017"
}

@article{Knapen:2016moh,
    author = "Knapen, Simon and Lin, Tongyan and Lou, Hou Keong and Melia, Tom",
    title = "{Searching for Axionlike Particles with Ultraperipheral Heavy-Ion Collisions}",
    eprint = "1607.06083",
    archivePrefix = "arXiv",
    primaryClass = "hep-ph",
    doi = "10.1103/PhysRevLett.118.171801",
    journal = "Phys. Rev. Lett.",
    volume = "118",
    number = "17",
    pages = "171801",
    year = "2017"
}

@article{Artamonov:2009sz,
    author = "Artamonov, A. V. and others",
    collaboration = "BNL-E949",
    title = "{Study of the decay $K^+\to\pi^+\nu \bar\nu$ in the momentum region $140 < P_\pi < 199$ MeV/c}",
    eprint = "0903.0030",
    archivePrefix = "arXiv",
    primaryClass = "hep-ex",
    reportNumber = "BNL-81786-2008-JA, FERMILAB-PUB-09-007-CD-T, KEK-2008-44, TRIUMF-TRI-PP-08-26, UHEP-EX-08-004",
    doi = "10.1103/PhysRevD.79.092004",
    journal = "Phys. Rev. D",
    volume = "79",
    pages = "092004",
    year = "2009"
}

@article{Fukuda:2015ana,
    author = "Fukuda, Hajime and Harigaya, Keisuke and Ibe, Masahiro and Yanagida, Tsutomu T.",
    title = "{Model of visible QCD axion}",
    eprint = "1504.06084",
    archivePrefix = "arXiv",
    primaryClass = "hep-ph",
    reportNumber = "IPMU15-0050",
    doi = "10.1103/PhysRevD.92.015021",
    journal = "Phys. Rev. D",
    volume = "92",
    number = "1",
    pages = "015021",
    year = "2015"
}

@article{Bauer:2018uxu,
    author = "Bauer, Martin and Heiles, Mathias and Neubert, Matthias and Thamm, Andrea",
    title = "{Axion-Like Particles at Future Colliders}",
    eprint = "1808.10323",
    archivePrefix = "arXiv",
    primaryClass = "hep-ph",
    reportNumber = "CERN-TH-2018-199, MITP/18-075",
    doi = "10.1140/epjc/s10052-019-6587-9",
    journal = "Eur. Phys. J. C",
    volume = "79",
    number = "1",
    pages = "74",
    year = "2019"
}

@article{Mariotti:2017vtv,
    author = "Mariotti, Alberto and Redigolo, Diego and Sala, Filippo and Tobioka, Kohsaku",
    title = "{New LHC bound on low-mass diphoton resonances}",
    eprint = "1710.01743",
    archivePrefix = "arXiv",
    primaryClass = "hep-ph",
    reportNumber = "DESY-17-148",
    doi = "10.1016/j.physletb.2018.06.039",
    journal = "Phys. Lett. B",
    volume = "783",
    pages = "13--18",
    year = "2018"
}

@article{CidVidal:2018blh,
    author = "Cid Vidal, Xabier and Mariotti, Alberto and Redigolo, Diego and Sala, Filippo and Tobioka, Kohsaku",
    title = "{New Axion Searches at Flavor Factories}",
    eprint = "1810.09452",
    archivePrefix = "arXiv",
    primaryClass = "hep-ph",
    reportNumber = "DESY-18-183",
    doi = "10.1007/JHEP01(2019)113",
    journal = "JHEP",
    volume = "01",
    pages = "113",
    year = "2019",
    note = "[Erratum: JHEP 06, 141 (2020)]"
}

@article{Aloni:2019ruo,
    author = "Aloni, Daniel and Fanelli, Cristiano and Soreq, Yotam and Williams, Mike",
    title = "{Photoproduction of Axionlike Particles}",
    eprint = "1903.03586",
    archivePrefix = "arXiv",
    primaryClass = "hep-ph",
    reportNumber = "CERN-TH-2019-023",
    doi = "10.1103/PhysRevLett.123.071801",
    journal = "Phys. Rev. Lett.",
    volume = "123",
    number = "7",
    pages = "071801",
    year = "2019"
}

@article{Sakaki:2020mqb,
    author = "Sakaki, Yasuhito and Ueda, Daiki",
    title = "{Searching for new light particles at the international linear collider main beam dump}",
    eprint = "2009.13790",
    archivePrefix = "arXiv",
    primaryClass = "hep-ph",
    doi = "10.1103/PhysRevD.103.035024",
    journal = "Phys. Rev. D",
    volume = "103",
    number = "3",
    pages = "035024",
    year = "2021"
}

@article{Florez:2021zoo,
    author = "Fl\'orez, Andr\'es and Gurrola, Alfredo and Johns, Will and Sheldon, Paul and Sheridan, Elijah and Sinha, Kuver and Soubasis, Brandon",
    title = "{Probing axionlike particles with $\gamma\gamma$ final states from vector boson fusion processes at the LHC}",
    eprint = "2101.11119",
    archivePrefix = "arXiv",
    primaryClass = "hep-ph",
    doi = "10.1103/PhysRevD.103.095001",
    journal = "Phys. Rev. D",
    volume = "103",
    number = "9",
    pages = "095001",
    year = "2021"
}

@article{Brdar:2020dpr,
    author = "Brdar, Vedran and Dutta, Bhaskar and Jang, Wooyoung and Kim, Doojin and Shoemaker, Ian M. and Tabrizi, Zahra and Thompson, Adrian and Yu, Jaehoon",
    title = "{Axionlike Particles at Future Neutrino Experiments: Closing the Cosmological Triangle}",
    eprint = "2011.07054",
    archivePrefix = "arXiv",
    primaryClass = "hep-ph",
    reportNumber = "FERMILAB-PUB-20-645-V, MI-TH-2029",
    doi = "10.1103/PhysRevLett.126.201801",
    journal = "Phys. Rev. Lett.",
    volume = "126",
    number = "20",
    pages = "201801",
    year = "2021"
}

@article{DallaValleGarcia:2023xhh,
    author = "Dalla Valle Garcia, Giovani and Kahlhoefer, Felix and Ovchynnikov, Maksym and Zaporozhchenko, Andrii",
    title = "{Phenomenology of axionlike particles with universal fermion couplings revisited}",
    eprint = "2310.03524",
    archivePrefix = "arXiv",
    primaryClass = "hep-ph",
    reportNumber = "TTP23-042, P3H-23-070",
    doi = "10.1103/PhysRevD.109.055042",
    journal = "Phys. Rev. D",
    volume = "109",
    number = "5",
    pages = "055042",
    year = "2024"
}

@article{Kyselov:2025uez,
    author = "Kyselov, Yehor and Mrenna, Stephen and Ovchynnikov, Maksym",
    title = "{New physics particles mixing with mesons: production in the fragmentation chain}",
    eprint = "2504.06828",
    archivePrefix = "arXiv",
    primaryClass = "hep-ph",
    reportNumber = "CERN-TH-2025-073, FERMILAB-PUB-25-0200-CSAID",
    month = "4",
    year = "2025"
}

@article{Afik:2023mhj,
    author = {Afik, Yoav and D\"obrich, Babette and Jerhot, Jan and Soreq, Yotam and Tobioka, Kohsaku},
    title = "{Probing long-lived axions at the KOTO experiment}",
    eprint = "2303.01521",
    archivePrefix = "arXiv",
    primaryClass = "hep-ph",
    reportNumber = "IRMP-CP3-23-11, IRMP-CP3-23-10, MPP-2023-40, KEK-TH-2499",
    doi = "10.1103/PhysRevD.108.055007",
    journal = "Phys. Rev. D",
    volume = "108",
    number = "5",
    pages = "055007",
    year = "2023"
}

@article{Balkin:2021jdr,
    author = "Balkin, Reuven and Krasny, Mieczyslaw W. and Ma, Teng and Safdi, Benjamin R. and Soreq, Yotam",
    title = "{Probing Axion-Like-Particles at the CERN Gamma Factory}",
    eprint = "2105.15072",
    archivePrefix = "arXiv",
    primaryClass = "hep-ph",
    doi = "10.1002/andp.202100222",
    journal = "Annalen Phys.",
    volume = "534",
    number = "3",
    pages = "2100222",
    year = "2022"
}

@article{Balkin:2023gya,
    author = "Balkin, Reuven and Hen, Or and Li, Wenliang and Liu, Hongkai and Ma, Teng and Soreq, Yotam and Williams, Mike",
    title = "{Probing axion-like particles at the Electron-Ion Collider}",
    eprint = "2310.08827",
    archivePrefix = "arXiv",
    primaryClass = "hep-ph",
    doi = "10.1007/JHEP02(2024)123",
    journal = "JHEP",
    volume = "02",
    pages = "123",
    year = "2024"
}

@article{Bai:2021gbm,
    author = "Bai, Zhaoyu and others",
    title = "{New physics searches with an optical dump at LUXE}",
    eprint = "2107.13554",
    archivePrefix = "arXiv",
    primaryClass = "hep-ph",
    reportNumber = "DESY 21-111",
    doi = "10.1103/PhysRevD.106.115034",
    journal = "Phys. Rev. D",
    volume = "106",
    number = "11",
    pages = "115034",
    year = "2022"
}

@article{Pybus:2023yex,
    author = "Pybus, J. R. and others",
    title = "{Search for axion-like particles through nuclear Primakoff production using the GlueX detector}",
    eprint = "2308.06339",
    archivePrefix = "arXiv",
    primaryClass = "hep-ex",
    doi = "10.1016/j.physletb.2024.138790",
    journal = "Phys. Lett. B",
    volume = "855",
    pages = "138790",
    year = "2024"
}

@article{Black:1998wt,
	author = "Black, Deirdre and Fariborz, Amir H. and Sannino, Francesco and Schechter, Joseph",
	title = "{Putative light scalar nonet}",
	eprint = "hep-ph/9808415",
	archivePrefix = "arXiv",
	reportNumber = "YCTP-21-98, SU-4240-683",
	doi = "10.1103/PhysRevD.59.074026",
	journal = "Phys. Rev. D",
	volume = "59",
	pages = "074026",
	year = "1999"
}

@article{Suzuki:1993zs,
    author = "Suzuki, M.",
    title = "{Tensor meson dominance: Phenomenology of the f2 meson}",
    doi = "10.1103/PhysRevD.47.1043",
    journal = "Phys. Rev. D",
    volume = "47",
    pages = "1043--1047",
    year = "1993"
}

@article{Han:1998sg,
    author = "Han, Tao and Lykken, Joseph D. and Zhang, Ren-Jie",
    title = "{On Kaluza-Klein states from large extra dimensions}",
    eprint = "hep-ph/9811350",
    archivePrefix = "arXiv",
    reportNumber = "MADPH-98-1092, FERMILAB-PUB-98-364",
    doi = "10.1103/PhysRevD.59.105006",
    journal = "Phys. Rev. D",
    volume = "59",
    pages = "105006",
    year = "1999"
}

@article{Katz:2005ir,
    author = "Katz, Emanuel and Lewandowski, Adam and Schwartz, Matthew D.",
    title = "{Tensor mesons in AdS/QCD}",
    eprint = "hep-ph/0510388",
    archivePrefix = "arXiv",
    reportNumber = "UCB-PTH-05-37, LBNL-59061, BUHEP-05-16",
    doi = "10.1103/PhysRevD.74.086004",
    journal = "Phys. Rev. D",
    volume = "74",
    pages = "086004",
    year = "2006"
}

\end{document}